\documentclass[a4paper,fleqn]{cas-sc}

\usepackage[authoryear]{natbib}

\usepackage{graphicx}
\usepackage{amssymb}
\usepackage{multirow}
\usepackage{pifont}
\usepackage{longtable}
\usepackage{bbold}
\usepackage{color}
\usepackage{lineno}

\makeatletter
    
    \newcommand{\Rmnum}[1]{\expandafter\@slowromancap\romannumeral #1@}
\makeatother

\def\tsc#1{\csdef{#1}{\textsc{\lowercase{#1}}\xspace}}
\tsc{WGM}
\tsc{QE}
\tsc{EP}
\tsc{PMS}
\tsc{BEC}
\tsc{DE}
\begin{document}
%\linenumbers
\let\WriteBookmarks\relax
\def\floatpagepagefraction{1}
\def\textpagefraction{.001}
\shorttitle{Hilda asteroids and free-floating planet flyby}
\shortauthors{Jian Li et~al.}

\title [mode = title]{Resonant amplitude distribution of the Hilda asteroids and the free-floating planet flyby scenario}                      
%\tnotemark[1,2]

\tnotetext[1]{This document is the results of the research
   project mainly funded by the National Natural Science Foundation of China.}

%\tnotetext[2]{The second title footnote which is a longer text matter
%   to fill through the whole text width and overflow into
%   another line in the footnotes area of the first page.}

\author[1,2]{Jian Li}[type=editor,
                        auid=000,bioid=1,
%                        prefix=Sir,
                        role=Corresponding author,
                        orcid=0009-0000-6832-4771]
\cormark[1]
\ead{ljian@nju.edu.cn}
%\ead[url]{www.jkkrishnan.in}

\credit{Conceptualization, Data curation, Formal Analysis, Funding acquisition, Methodology, Investigation, Software, Writing – original draft, Writing – review \& editing}

%\address[1]{, Street 129, 1043 NX Amsterdam, The Netherlands}
\affiliation[1]{organization={School of Astronomy and Space Science, Nanjing University},
                addressline={163 Xianlin Avenue}, 
                city={Nanjing},
%               citysep={}, % Uncomment if no comma needed between city and postcode
                postcode={210023}, 
%                state={Kerala},
                country={PR China}}

\affiliation[2]{organization={Key Laboratory of Modern Astronomy and Astrophysics in Ministry of Education, Nanjing University},
                addressline={163 Xianlin Avenue}, 
                city={Nanjing},
%               citysep={}, % Uncomment if no comma needed between city and postcode
                postcode={210023}, 
%                state={Kerala},
                country={PR China}}

\author[3,4]{Zhihong Jeff Xia}[%
   role=Co-corresponding author,
%   suffix=Jr,
   ]
\ead{xia@math.northwestern.edu}

\credit{Conceptualization, Methodology, Investigation, Writing – original draft, Writing – review \& editing}

\affiliation[3]{organization={School of Sciences, Great Bay University},
                addressline={16 University Road, Songshan Lake}, 
                city={Dongguan},
                postcode={523808}, 
%                state={Orissa}, 
                country={PR China}}

\affiliation[4]{organization={Department of Mathematics, Northwestern University},
                addressline={2033 Sheridan Road}, 
                city={Evanston},
                postcode={60208}, 
                state={Illinois}, 
                country={USA}}

\author[1,2]{Hanlun Lei}
%\ead[URL]{https://www.university.org}
\credit{Methodology, Investigation, Funding acquisition, Writing – original draft, Writing – review \& editing}

\author[5,6]{Nikolaos Georgakarakos}

\credit{Investigation, Writing – original draft, Writing – review \& editing}

\affiliation[5]{organization={New York University Abu Dhabi},
%                addressline={16 University Road, Songshan Lake}, 
                city={Abu Dhabi},
                postcode={PO Box 129188}, 
%                state={Orissa}, 
                country={United Arab Emirates}}

\affiliation[6]{organization={Center for Astrophysics and Space Science (CASS), New York University Abu Dhabi},
%                addressline={16 University Road, Songshan Lake}, 
                city={Abu Dhabi},
                postcode={PO Box 129188}, 
%                state={Orissa}, 
                country={United Arab Emirates}}

\author[7,8]{Fumi Yoshida}

\credit{Investigation, Funding acquisition, Writing – review \& editing}

\affiliation[7]{organization={University of Occupational and Environmental Health},
                addressline={1-1 Iseigaoka, Yahata}, 
                city={Kitakyusyu},
                postcode={807-8555}, 
%                state={Orissa}, 
                country={Japan}}

\affiliation[8]{organization={Planetary Exploration Research Center, Chiba Institute of Technology},
                addressline={1-1 Iseigaoka, Yahata}, 
%                addressline={2-17-1 Tsudanuma, Narashino}, 
                city={Chiba},
                postcode={275-0016}, 
%                state={Orissa}, 
                country={Japan}}
                
%\cormark[2]
%\fnmark[1,3]
%\ead{t.rafeeq@example.in}
%\ead[URL]{www.campus.in}

\author[9]{Xin Li}

\credit{Investigation, Writing – review \& editing}

\affiliation[9]{organization={Department of Computer Science, Northwestern University},
                addressline={2233 Tech Drive}, 
                city={Evanston},
                postcode={60208}, 
                state={Illinois}, 
                country={USA}}

\cortext[cor1]{Jian Li}
%\cortext[cor2]{Principal corresponding author}
%\fntext[fn1]{This is the first author footnote, but is common to third
%  author as well.}
%\fntext[fn2]{Another author footnote, this is a very long footnote and
%  it should be a really long footnote. But this footnote is not yet
%  sufficiently long enough to make two lines of footnote text.}

%\nonumnote{This note has no numbers. In this work we demonstrate $a_b$
%  the formation Y\_1 of a new type of polariton on the interface
%  between a cuprous oxide slab and a polystyrene micro-sphere placed
%  on the slab.
%  }

\begin{abstract}
In some recent work, we provided a quantitative explanation for the number asymmetry of Jupiter Trojans by hypothesizing a free-floating planet (FFP) flyby into the Solar System. In support of that explanation, this paper examines the influence of the same FFP flyby on the Hilda asteroids, which orbit stably in the 3:2 mean motion resonance with Jupiter. The observed Hilda population exhibits two distinct resonant patterns: (1) a lack of Hildas with resonant amplitudes $<40^{\circ}$ at eccentricities $<0.1$; (2) a nearly complete absence of Hildas with amplitudes $<20^{\circ}$, regardless of eccentricity. Previous models of Jupiter migration and resonance capture could account for the eccentricity distribution of Hildas but have failed to replicate the unusual absence of those with the smallest resonant amplitudes, which theoretically should be the most stable. Here we report that the FFP flyby can trigger an extremely rapid outward migration of Jupiter, causing a sudden shift in the 3:2 Jovian resonance. Consequently, Hildas with varying eccentricities would have their resonant amplitudes changed by different degrees, leading to the observed resonant patterns. We additionally show that, in our FFP flyby scenario, these patterns are consistently present across different resonant amplitude distributions of primordial Hildas arising from various formation models. We also place constraints on the potential parameters of the FFP, suggesting it should have an eccentricity of 1-1.3 or larger, an inclination up to $30^{\circ}$ or higher, and a minimum mass of about 50 Earth masses.
\end{abstract}

%\includegraphics{figs/cas-grabs.pdf}

%\item A sudden outward jump of Jupiter can simultaneously explain both the number asymmetry of Jupiter Trojans and the resonance amplitude distribution of Hildas.

%\item The reproduction of the observed resonant patterns of Hildas is nearly independent of their various formation models

%\item The proposed jumping Jupiter can potentially be caused by a free-floating planet flyby, which may reshape the entire architecture of our Solar System.

\begin{keywords}
Solar system evolution \sep Solar system planets \sep Jupiter \sep Hilda group \sep Dynamical evolution \sep Free floating planets
\end{keywords}
 
\maketitle

\section{Introduction}

The hypothesis of the free-floating planet (FFP) flyby originated from the number asymmetry of L4 and L5 Jupiter Trojans, a long-standing mystery dating back over 30 years \citep{shoe89, jewi04, frei06, grav11, grav12, li2018}. This number discrepancy puts forward a great challenge to planetary evolution theories, prompting the proposal of various dynamical models to investigate the origin of Jupiter Trojans \citep[e.g.][]{Nesv13, pira19, deie22, Li2023a}. In our recent study \citep{Li2023}, we demonstrated that if an FFP with sub-Saturn to Jupiter mass flew by the Solar System, the orbit of Jupiter could rapidly expand by tenths of an AU over a short timescale of 10 yr. This sudden outward migration of Jupiter asymmetrically distorted the phase space near the L4 and L5 triangular Lagrangian points of the Sun-Jupiter system, resulting in an L4-to-L5 number ratio of $\sim 1.6$ for the Jupiter Trojans, providing a quantitative explanation for the current unbiased observation \citep{szab07}. Such rapid migration of Jupiter appears too fast to occur during the early dynamical instability stage of planet-planet scattering within the Solar System \citep{Tsig05,Nesv13}. Moreover, the introduction of an FFP can naturally trigger Jupiter to perform an outward jump, whereas in the planet instability scenario, Jupiter generally needs to jump inwards once it ejects the fifth giant planet \citep{Li2023a}.

By now, two interstellar objects, 1I/`Oumuamua \citep{Will17,Meec17} and 2I/Borisov \citep{Toma19,Guzi19,Guzi20} have been discovered passing through the Solar System. Their high incoming velocities suggest that they originated outside our planetary system \citep{Raym18,Jack18,Hand19,Bail20,Hall20}. This discovery raises the possibility that a more massive interstellar celestial body, such as an FFP, could also visit us for a brief period. This is possible, as FFPs are abundant throughout the Milky Way and beyond into the extragalactic regime \citep{Sumi11,Bhat19,Sumi23}. 
%If a massive FFP were to enter a planetary system, it may significantly reshape its architecture, leaving discernible imprints behind. For instance, concerning extrasolar systems, \citet{Moor23} proposed that a recent close encounter with an FFP could account for the observed structure of HD 106906, characterized by a warped debris disk and an inclined companion. 
In theory, \citet{Doul19} conducted a systematic study on the interaction between an FFP and an existing planetary system. This study considered two scenarios: a Solar-like system consisting of the Sun and Jupiter at 5.2 AU, and an extrasolar system with a Jupiter-sized planet orbiting a Sun-like star at 1 AU.
In practical observation, if a massive FFP were to enter a planetary system, it may significantly reshape its architecture, leaving discernible imprints behind. For instance, concerning extrasolar systems, \citet{Moor23} proposed that a recent close encounter with an FFP could account for the observed structure of HD 106906, characterized by a warped debris disk and an inclined companion.

To support our previously published hypothesis that the number asymmetry of Jupiter Trojans can be explained by an FFP flyby, we intend to seek other dynamical signatures within the Solar System. In the same FFP flyby scenario, other existing asteroid populations could also be affected, though the resulting structures must not conflict with current observations. We recall that the key mechanism in this scenario is the outward jump of Jupiter triggered by the FFP. As the Jovian mean motion resonances (MMRs) move along with Jupiter, after considering the 1:1 Jovian MMR where Jupiter Trojans are settled, the next asteroid population that naturally comes to mind is the Hilda asteroids. The Hildas reside in the 3:2 Jovian MMR, and a significant number of them have been observed. We speculate that the FFP may also have left a signature in the Hildas, which could provide a useful test of the validity of our FFP hypothesis regarding Jupiter Trojans. Notably, the Jupiter Trojan and Hilda populations are found to share remarkable similarities in physical properties and thus may have a common origin \citep{Wong18}.

\begin{figure}
  \centering
  \begin{minipage}[c]{0.5\textwidth}
  \centering
  \hspace{0cm}
  \includegraphics[width=9cm]{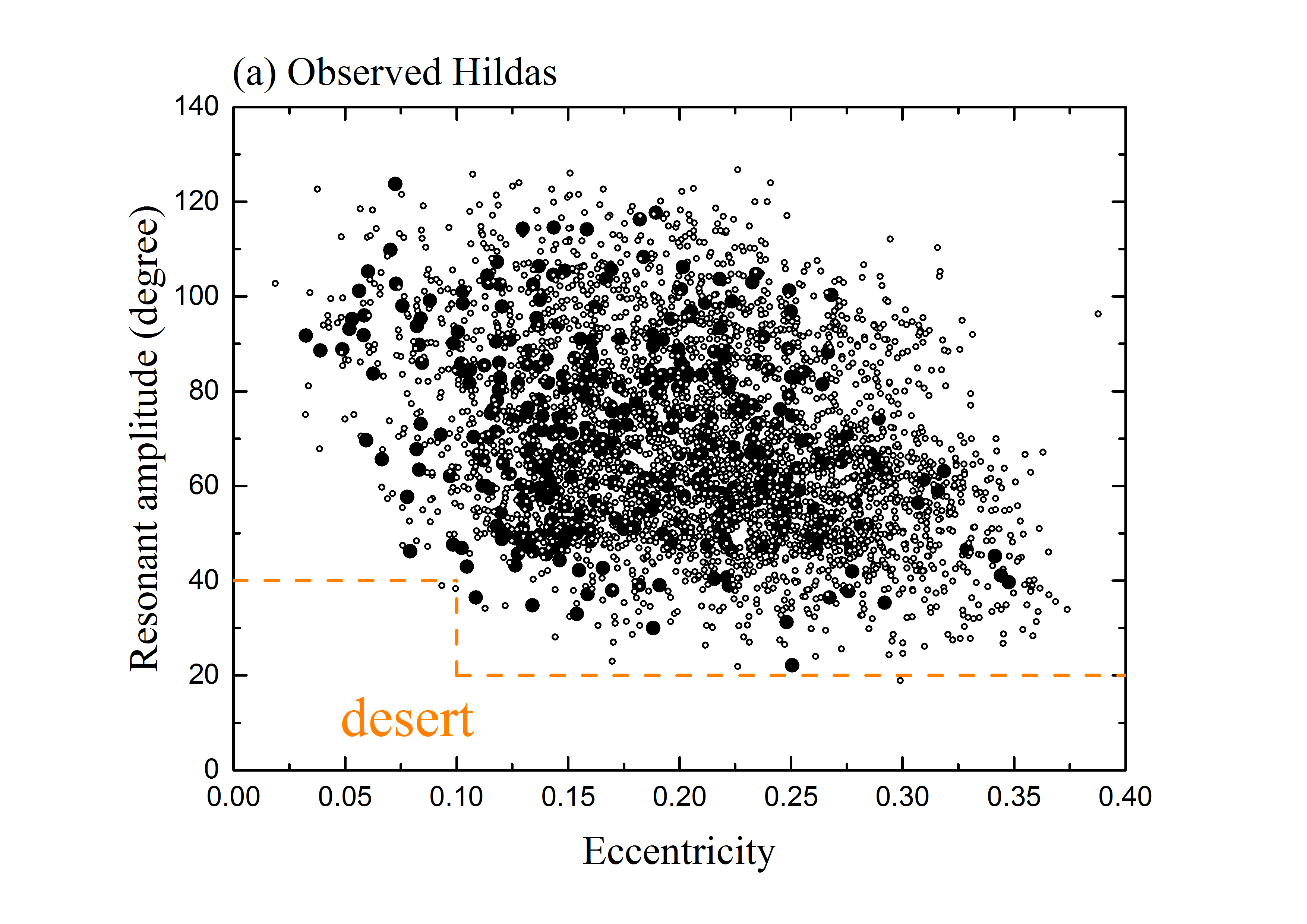}
  \end{minipage}
  \begin{minipage}[c]{0.5\textwidth}
  \centering
  \hspace{0cm}
  \includegraphics[width=9cm]{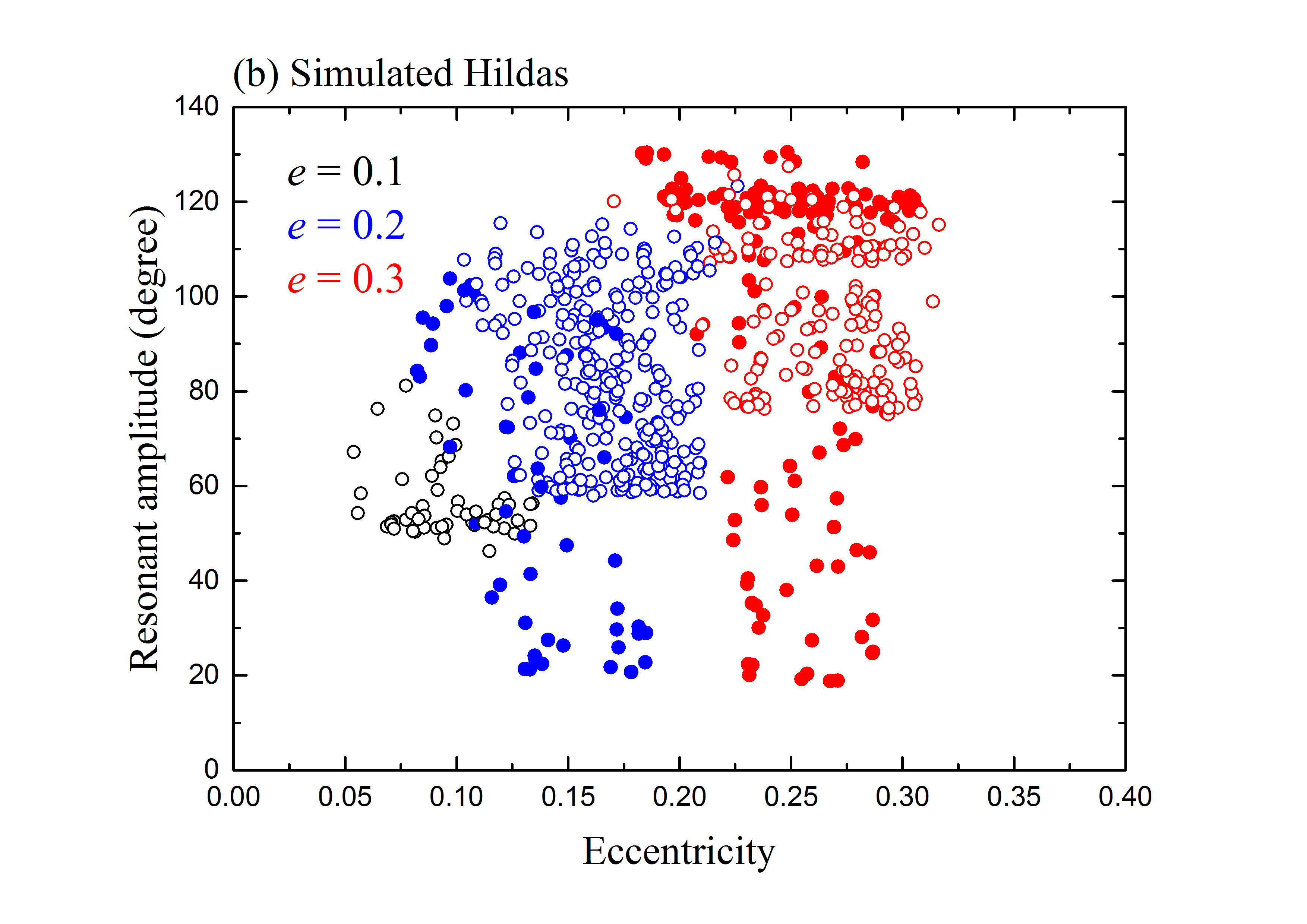}
  \end{minipage}
 \caption{Distribution of eccentricities ($e$) and resonant amplitudes ($A$) of the observed and simulated Hildas. (Panel a) The observed Hildas, as of 2020 July, consist of over 3700 objects recorded in the IAU Minor Planet Center. Their resonant amplitude distribution exhibits two distinct patterns: a lack of objects with $A<40^{\circ}$ at $e\lesssim 0.1$, and nearly none with $A<20^{\circ}$. The large dots indicate the objects with diameters $>$ 10 km (equivalent to absolute magnitudes $<$ 13.8, assuming a typical albedo of 0.05). For reference, two horizontal dashed lines are plotted at $A=20^{\circ}$ and $40^{\circ}$ respectively and a vertical dashed line is plotted at $e=0.1$. (Panel b) The simulated Hildas, which start on Hilda-like orbits with $e=0.1$ (black), 0.2 (blue), and 0.3 (red), survive in the Jovian 3:2 resonance at the end of the 1 Myr evolution in the FFP flyby model. The dots stand for Population \Rmnum1, initially located at the nominal resonance with the same semi-major axis $a=a_{res}=3.97$ AU, but with different resonant angles $\sigma=-150^{\circ}$-$150^{\circ}$. The circles stand for Population \Rmnum2, initially settled at the resonant centre of $\sigma=0^{\circ}$, but with varying $a$.}
 \label{all}
\end{figure}

Currently, more than 3700 Hilda asteroids have been registered in the IAU Minor Planet Center (MPC)\footnote{https://www.minorplanetcenter.net/iau/MPCORB}. The identification process starts by selecting the asteroids with semi-major axes $a=3.8$-4.1 AU from the MPC Orbit (MPCORB) Database and only the objects with observations at two or more oppositions are considered. This chosen $a$-range encompasses the nominal location of the 3:2 Jovian resonance, which is approximately at $\sim3.97$ AU. Next, to identify the Hilda asteroids, we numerically integrate the orbits of these observed asteroids for a time-span of 1 Myr, under the gravitational perturbations of the four giant planets in the outer Solar System. As for the planets, their masses, initial positions, and velocities are adopted from DE441 \citep{Park21}, in the heliocentric frame referred to the J2000.0 ecliptic plane. The epoch of the state vectors of the planets is chosen to be 2020 May 31, and it is the same as that of the asteroids. Finally, we examine the time evolution of the resonant angle $\sigma=2\lambda-3\lambda_J+\varpi$, where $\lambda$ and $\varpi$ are the mean longitude and the longitude of perihelion of the asteroid, respectively, while $\lambda_J$ is the mean longitude of Jupiter. Then an asteroid is identified as a Hilda only if it can consistently exhibit libration of $\sigma$, equivalent to requiring a resonant amplitude of $A<180^{\circ}$, throughout the 1 Myr evolution.

Figure \ref{all} (a) displays the distribution of eccentricities and resonant amplitudes of the observed Hildas. Two distinct patterns are easily observed: there are no objects with resonant amplitudes $A<40^{\circ}$ in the low eccentricity region of $e<0.1$ and almost no asteroids on orbits with $A<20^{\circ}$. Such desert in the resonant amplitude distribution seems unusual because the Hildas with small resonant amplitudes reside deeply in the 3:2 resonance and should be very stable. Previous models \citep{Fran04,Broz11} of Jupiter migration and resonance capture could account for their eccentricity distribution but cannot replicate the absence of the Hildas with the smallest resonant amplitudes. Actually, the accumulation of Hildas towards large $A$ has been noticed before and attributed to the fact that objects with low $e$ generally have larger $A$ than those with high $e$ \citep{Fran04,Schu82,Schu91}. However, the reason why low-$e$ Hildas tend to have larger $A$ remains unexplained. Furthermore, even when the Yarkovsky effect is taken into account, the Hildas with diameters exceeding 10 km are little affected \citep{Broz08}. As indicated by the large dots in Fig. \ref{all} (a), the overall $A$-distribution in this size range resembles that of the entire Hilda population. Therefore, we conclude that the observed desert in the smallest-$A$ region of the Hildas should be reliable.

%This paper aims to test the validity of the FFP flyby hypothesis regarding Jupiter Trojans proposed by \citet{Li2023} by exploring whether the same FFP flyby can produce the distinct patterns of the Hilda desert similar to those inferred from Figure \ref{all}(a). The rest of this paper is organized as follows. In Sect. 2, within the framework of the FFP flyby and the resulting Jupiter jump, we investigate the evolution of Hildas starting from a uniform distribution of resonant amplitudes. Remarkably, we find that the resonant amplitude distribution of the simulated Hildas aligns well with observations. In Sect. 3, we further verify our FFP hypothesis for replicating the resonant patterns of the Hildas. Section 4 places constraints on the parameters of the potential FFP flyby. The conclusions and discussion are given in Sect. 5.

This paper aims to test the validity of the FFP flyby hypothesis regarding Jupiter Trojans proposed by \citet{Li2023} by exploring whether the same FFP flyby can produce the distinct patterns of the Hilda desert similar to those inferred from Figure \ref{all} (a). The rest of this paper is organized as follows. In Sect. 2, within the framework of the FFP flyby and the resulting Jupiter jump, we numerically investigate the evolution of Hildas starting from a uniform distribution of resonant amplitudes. Remarkably, we find that the resonant amplitude distribution of the simulated Hildas aligns well with observations. 
And, other possible initial distributions of resonant amplitudes of Hildas have also been discussed. In Sect. 3, we construct a theoretical approach to demonstrate how our FFP hypothesis could replicate the resonant patterns of the Hildas.
Section 4 places constraints on the parameters of the potential FFP flyby. The conclusions and discussion are given in Sect. 5.

%_____________________________________________________________________________________________________________________

\section{Model and numerical results}

\subsection{FFP flyby and Jupiter jump}

We employ the same FFP flyby model that was used to explain the L4-to-L5 number asymmetry of $\sim1.6$ for the observed Jupiter Trojans in our previous work \citep{Li2023}. This FFP flyby could cause Jupiter to migrate outwards by $\sim0.12$ AU on a timescale of about 10 yr. Within such a dynamical model, we explore the evolution of the Hildas through numerical simulations, focusing on their resonant amplitude variations.

First, we give a brief overview of the aforementioned FFP flyby model. We utilise a coplanar three-body system consisting of the Sun, Jupiter, and an incoming FFP. In the heliocentric framework, Jupiter is set to initially move on a Keplerian circular orbit with a semi-major axis of $a_J=a_J^{\ast}=5.2$ AU and a phase angle of $\theta=255^{\circ}$ (anticlockwise from the positive $x$-axis), while the FFP is introduced on a parabolic trajectory with initial cartesian coordinates of $(x=-40a_J^{\ast}, y=-13a_J^{\ast})$ (see Fig. 1 in \citet{Li2023}). This ensures that the FFP has a perihelion that is about 0.15 AU beyond Jupiter's orbit, resulting in a close encounter but not a collision. 
%We note that, from here and below, the Solar System models are theoretical and no actual ephemerides are used. 
As for the mass of the FFP, it is adopted to be $m_{FFP}=m_J$, where $m_J$ represents the mass of Jupiter.

Around the time of 250 yr, the close encounter between Jupiter and the FFP takes place. Consequently, Jupiter has its motion accelerated and migrates outwards by $\Delta a_J\sim0.12$ AU on a timescale of 10 yr, while the FFP travels far away from the Solar System soon after and it stops being considered anymore in our model. This specific migration magnitude of Jupiter, i.e. $\Delta a_J\sim0.12$ AU, can  quantitatively reproduce the number asymmetry of Jupiter Trojans. It's worth noting that to achieve the desired $\Delta a_J$ value, the FFP is allowed to possess various parameters. For simplicity, we adopt a parabolic orbit for the FFP in this study, as we did in \citet{Li2023}, while the interstellar FFP passing through the Solar System was clearly hyperbolic. In Sect. \ref{orbitandmass} of this paper, we will conduct a series of FFP flyby models to deduce potential parameters for the FFP, such as eccentricity, inclination, and mass.

\begin{figure*}
  \centering
  \begin{minipage}[c]{1\textwidth}
  \vspace{0 cm}
  \includegraphics[width=9cm]{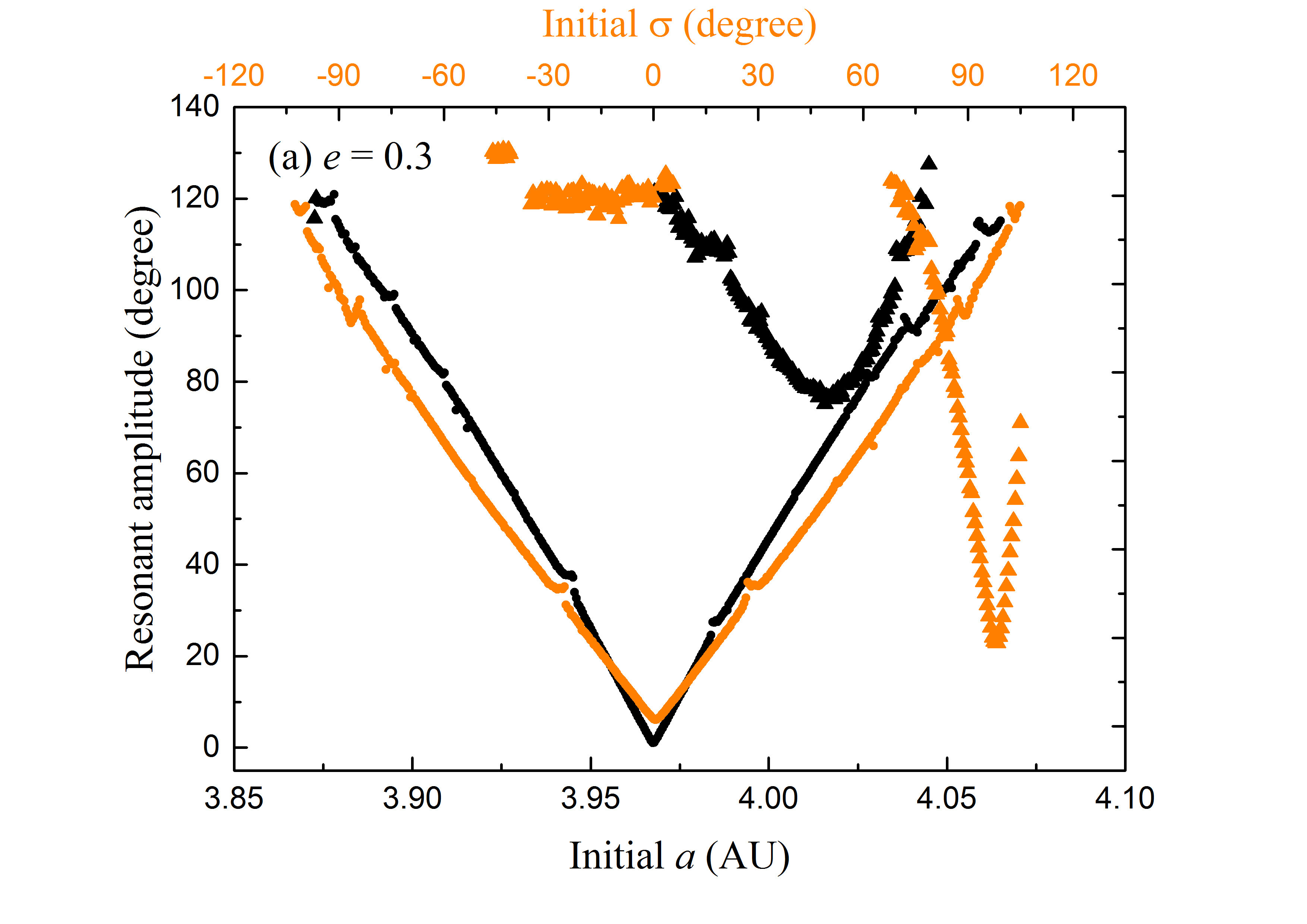}
  \end{minipage}
  \begin{minipage}[c]{1\textwidth}
  \vspace{0 cm}
  \includegraphics[width=9cm]{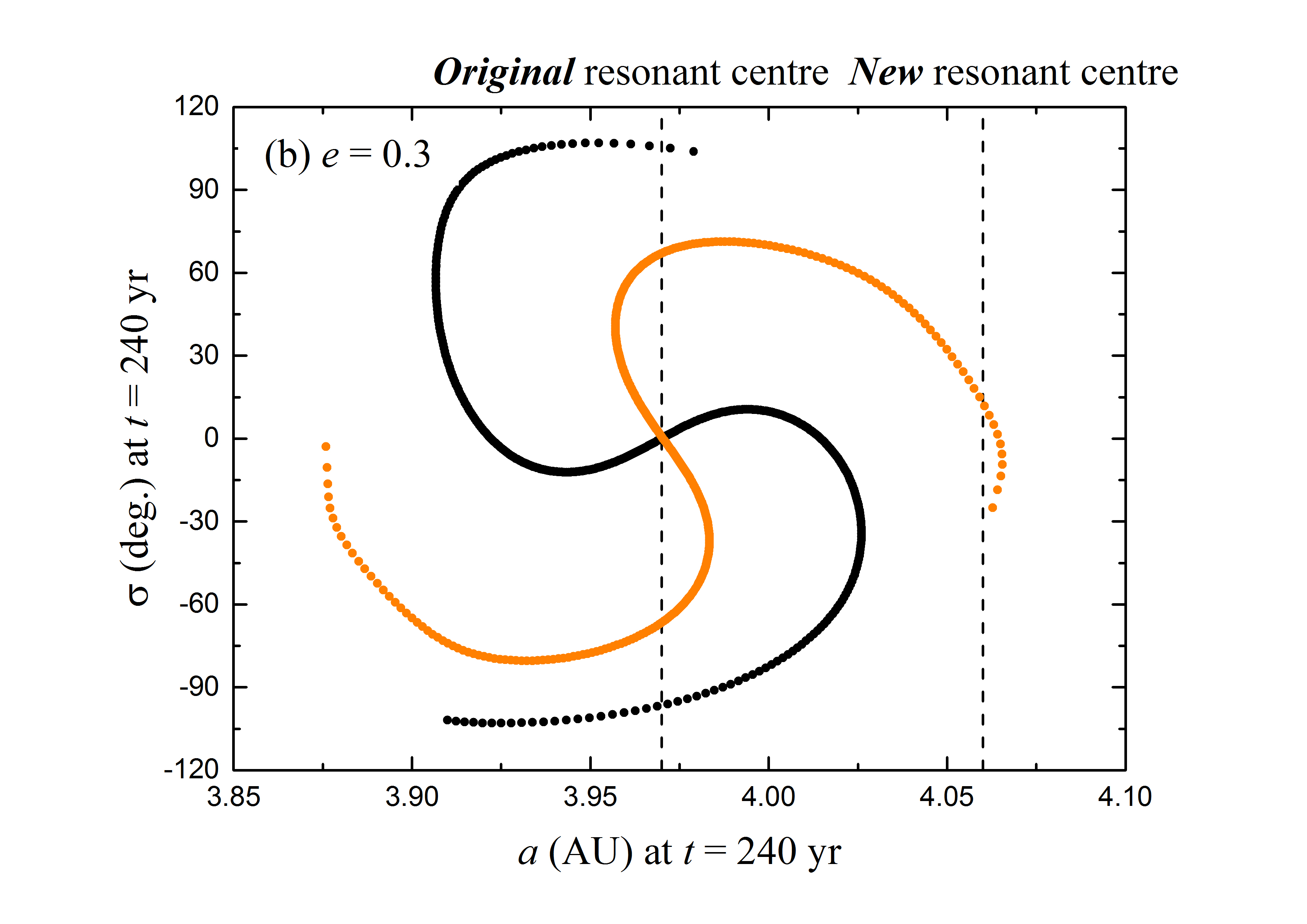}
  \includegraphics[width=9cm]{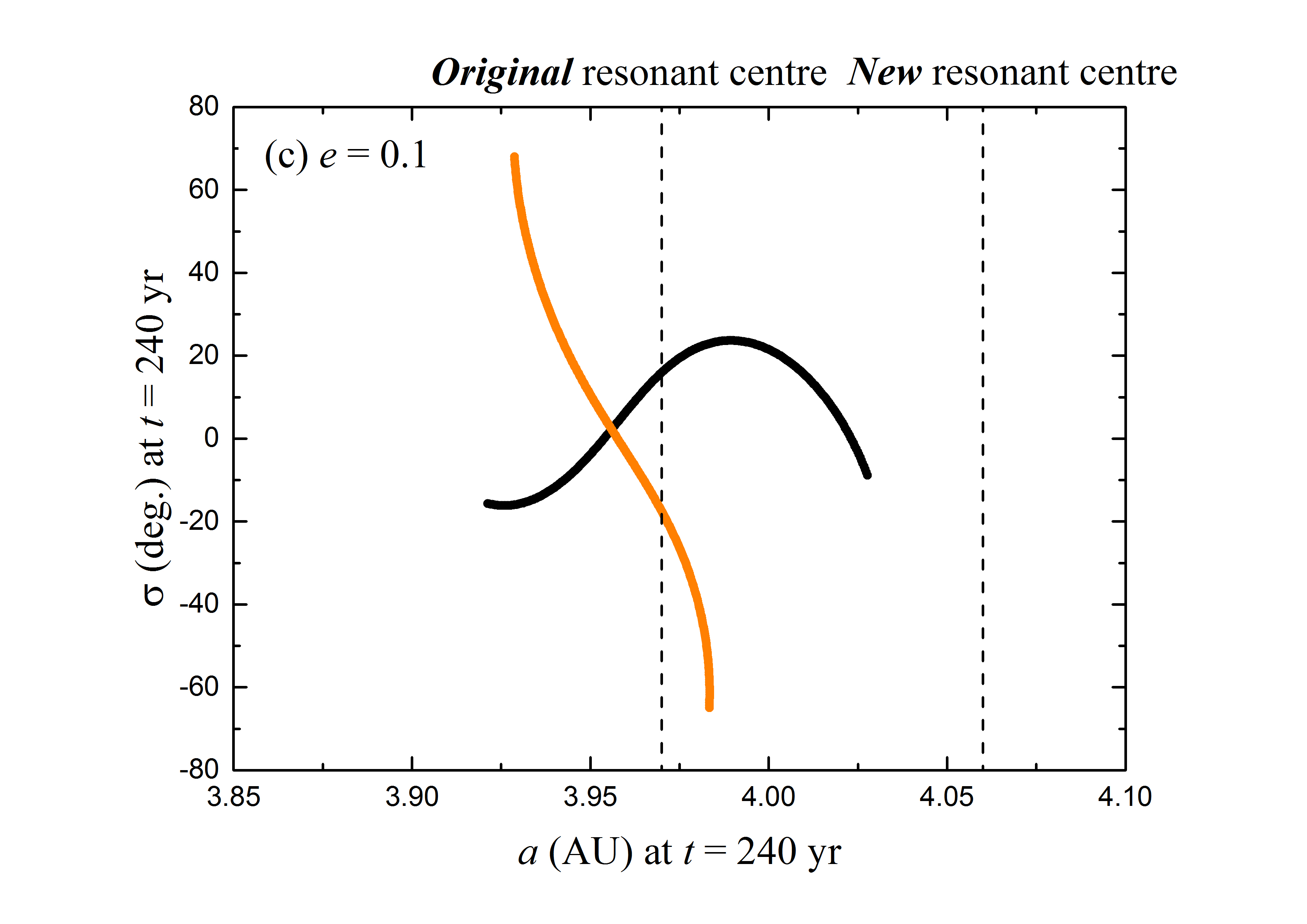}
  \end{minipage}
   \vspace{0 cm}
  \caption{Resonant dynamics of the simulated Hildas in the scenario with (triangles) and without (dots) the flyby of an FFP. The orange and black symbols refer to Population \Rmnum1 and Population \Rmnum2, respectively. (Panel a) For the case of $e=0.3$, the resonant amplitude distribution of test Hildas at the end of the 1 Myr evolution. (Panel b) Also for the case of $e=0.3$, the $a$-$\sigma$ distribution of test Hildas at $t=240$ yr. Shortly after this time, the FFP approaches its perihelion and drives Jupiter to jump outwards, causing the resonant centre to move from the original location ($a\sim 3.97$ AU) to the new one ($a\sim 4.06$ AU), as indicated by the two dashed vertical lines. Note that a small fraction of test Hildas can touch the new resonant centre. When the FFP flyby is involved, they could become the objects with $A\sim20^{\circ}$ as indicated by the lowest orange triangles in panel (a). (Panel c) Similar to panel (b) but for $e=0.1$, where none of the test Hildas can be found near the new resonant centre.}
 \label{amplitude}
\end{figure*} 

\subsection{Generation and evolution of test Hildas}

Regarding the test Hildas, we initially distribute them into two typical structures in the resonant phase space \citep{Li2022}: (1) Population \Rmnum1, which consists of 500 particles starting with a fixed semi-major axes of $a=a_{res}=3.97$ AU corresponding to the current nominal position of the 3:2 MMR and with resonant angles $\sigma$ ranging from $-150^{\circ}$ to $150^{\circ}$; (2) Population \Rmnum2, which includes 600 particles starting with fixed $\sigma=0^{\circ}$ corresponding to the 3:2 resonant centre and with varying $a$ values. Theoretically speaking, all other $a$-$\sigma$ distributions in the resonant phase space fall between these two options. The eccentricities of test Hildas are chosen to have three representative values of $e=0.1$, 0.2, and 0.3.  We also assume that the orbits of all bodies lie in the same plane. This is reasonable as most Hildas have small inclinations of $i<10^{\circ}$; in addition, $i$ is not a significant factor in the considered 3:2 MMR, which is of the eccentricity-type.

We first conduct pre-runs lasting 1 Myr for the test Hildas from Populations \Rmnum1 and \Rmnum2. In our pre-runs, the FFP is not included. This allows us to extract the consistently librating objects, which are then selected as the initial samples in the FFP flyby model. By doing so, we can ensure that the selected test Hildas are relatively stable and can well mimic the in-situ Hildas before the FFP approaches the Solar System.

Then, in the model of the FFP flyby and jumping Jupiter, we carry out numerical simulations to track the evolution of stable test Hildas from pre-runs. We start by performing a 300 yr simulation for observing the evolution of the planets and resonant behaviours of test Hildas. To accurately integrate the orbital evolution of these bodies during the violent stage of the close encounter between the FFP and Jupiter, which occurs around 250 yr as mentioned above, we use the 7-8th-order Runge-Kutta-Fehlberg (RKF) algorithm with an adaptive step-size and a local truncation error of less than $10^{-15}$. At the end of the 300 yr simulation, the FFP has gone far enough and only Jupiter and test Hildas remain in our system. Afterward, the evolution of Hildas, which would be much more regular under the perturbation of just one planet (Jupiter), is extended for another 1 Myr to account for their stability. To fulfill the orbital integration within a reasonable time, we employ the SWIFT\_RMVS3 symplectic integrator with a time step of 0.5 yr, which is about 1/16 of the shortest orbital period (Hildas) in the considered system \citep{Levi94}. The resonant survivors in the 3:2 Jovian resonance at the end of the 1 Myr simulation are regarded to be simulated Hildas, as shown in Fig. \ref{all} (b).

\subsection{Preliminary results}
\label{mechanism} 

In Fig. \ref{all} (b), we present the distribution of the simulated Hildas after 1 Myr evolution due to the flyby of an FFP. We find that the objects with small eccentricities of $e\lesssim 0.1$ possess quite large resonant amplitudes of $A>40^{\circ}$ and nearly all objects, regardless of their eccentricities, have resonant amplitudes of $A>20^{\circ}$. Such distinct patterns perfectly match what we see in the observed Hildas shown in Fig. \ref{all} (a). This suggests that the FFP flyby scenario could provide a quantitative agreement in the resonant amplitude distribution between observations and simulations. Then, it is of great interest to understand the underlying mechanism behind the numerical results by analysing the dynamics of the 3:2 Jovian MMR in the context of the FFP flyby.

We first consider the case of test Hildas starting with $e=0.3$, as displayed in Fig. \ref{amplitude} (a). Without the FFP flyby, their resonant amplitudes are symmetrically distributed around the resonant centres of $\sigma=0^{\circ}$ for Population \Rmnum1 (orange dots) and $a\approx3.97$ AU for Population \Rmnum2 (black dots). These $\sigma$ and $a$ values correspond to the original resonant centres, as expected in the Sun+Jupiter+Hilda restricted 3-body model. However, when the FFP flyby is taken into account, the profile of the resonant amplitude distribution generated from the same test Hilda samples could be significantly different. For Population \Rmnum1 (see orange triangles), the test Hildas initially close to the original resonant centre of $\sigma=0^{\circ}$ end up with very large resonant amplitudes of $A\sim120^{\circ}$, while those far away with initial $\sigma>60^{\circ}$ have smaller $A$. In particular, the objects starting with $\sigma\sim90^{\circ}$ evolve to orbits with $A$ as low as $20^{\circ}$ due to the inclusion of the FFP. Actually, the initial $a\sim4.06$ AU of these small-$A$ survivors refers to the new resonant centre discussed immediately after. 

To understand these changes, we outline the major relevant events that happen along the timeline. At the beginning of the integration, the FFP enters the Solar System on a parabolic orbit from 200 AU away. At time $t=240$ yr, we extract the $a$-$\sigma$ distribution of stably librating test Hildas from the case without FFP, as plotted in Fig. \ref{amplitude} (b). By this time, the influence of the FFP has not taken place yet and these objects are actually the candidate Hildas to be considered in the FFP flyby model. It should be noted that the typical period of the 3:2 Jovian resonance is of the order of 200-300 yr and therefore, the resonant phases of test Hildas at $t=240$ yr are substantially different from their initial conditions. Shortly after, around $t=250$ yr, the FFP passes through its perihelion and approaches the orbit of Jupiter, causing Jupiter to migrate outwards by $\sim0.12$ AU. Accordingly, the 3:2 Jovian resonance experiences radial displacement as its resonant centre jumps from the original location at $a\sim3.97$ AU to the new one at $a\sim4.06$ AU. As shown in Fig. \ref{amplitude} (b), there is a fraction of Population \Rmnum1 objects (in orange) touching the new resonant centre. We then realize that these objects can survive deeply in the new 3:2 Jovian resonance after the event of the FFP flyby and Jupiter migration, eventually evolving into the Hildas with $A$ as small as $\sim20^{\circ}$, which are represented by those lowest orange triangles on the right side of Fig. \ref{amplitude} (a).

The evolution of test Hildas with initial $e=0.2$ closely resembles that of the  case of $e=0.3$ presented above. However, in the case of $e=0.1$, Fig. \ref{amplitude} (c) shows that there are no test Hildas located near the new resonant centre at $a\sim4.06$ AU. This naturally explains the lack of small eccentricity Hildas at $A<40^{\circ}$ as seen in Fig. \ref{all} (b). 

We would like to mention that, to focus on Jupiter's 3:2 resonance, we have neglected the effects of the other three giant planets in our simulations for the study of test Hildas. This is justifiable because previous studies have shown that, for example, the closest planet, Saturn, has little impact on the overall distributions of the observed Hilda asteroids in eccentricity, inclination, and resonant amplitude \citep[e.g.][]{Schu91}. Furthermore, \citet{Schu07} investigated the three-body resonance for Hildas by considering the combined action of Jupiter and Saturn, and has demonstrated that the 3:2 resonant angles $\sigma$ may transition from libration (i.e. $A<180^{\circ}$) to circulation (i.e. $A=180^{\circ}$) and vice versa. Nevertheless, this mechanism mainly affects the particular Hilda asteroids that are close to the three-body resonance, while the majority of them are located far away from this resonance. Especially, the Hilda population with $A<60^{\circ}$ would not be influenced at all, indicating that if there were Hildas with $e=0.1$ having smaller $A<40^{\circ}$, their orbits would remain unaffected by the three-body resonance. As we will verify in the next section, although the secular perturbation of Saturn, Uranus, and Neptune could somewhat excite the resonant amplitudes of the Hildas, our FFP flyby scenario may still need to be involved to replicate the observed patterns. 
%Inside Jupiter's 3:2 resonance region, there are also other relevant resonances could be at play. \citet{Henr96} has displayed the localization of the secondary resonances inside this 3:2 resonance in the dynamical map on $(a, e)$ plane. Actually, such secondary resonances mainly occupy a very narrow band at $e<0.1$, so they should not be directly relevant to this work.
Inside Jupiter's 3:2 resonance region, there are also other relevant resonances that could be at play. \citet{Henr96} displayed the localization of the secondary resonances inside this 3:2 resonance on the dynamical map in the $(a, e)$ plane. Actually, these secondary resonances are very close to the separatrices of the 3:2 resonances, so they would not have a direct impact on the evolution of the Hildas near the resonance centre considered in this work.

\subsection{Further verification}

\begin{figure*}
  \centering
  \begin{minipage}[c]{1\textwidth}
  \vspace{0 cm}
  \includegraphics[width=9cm]{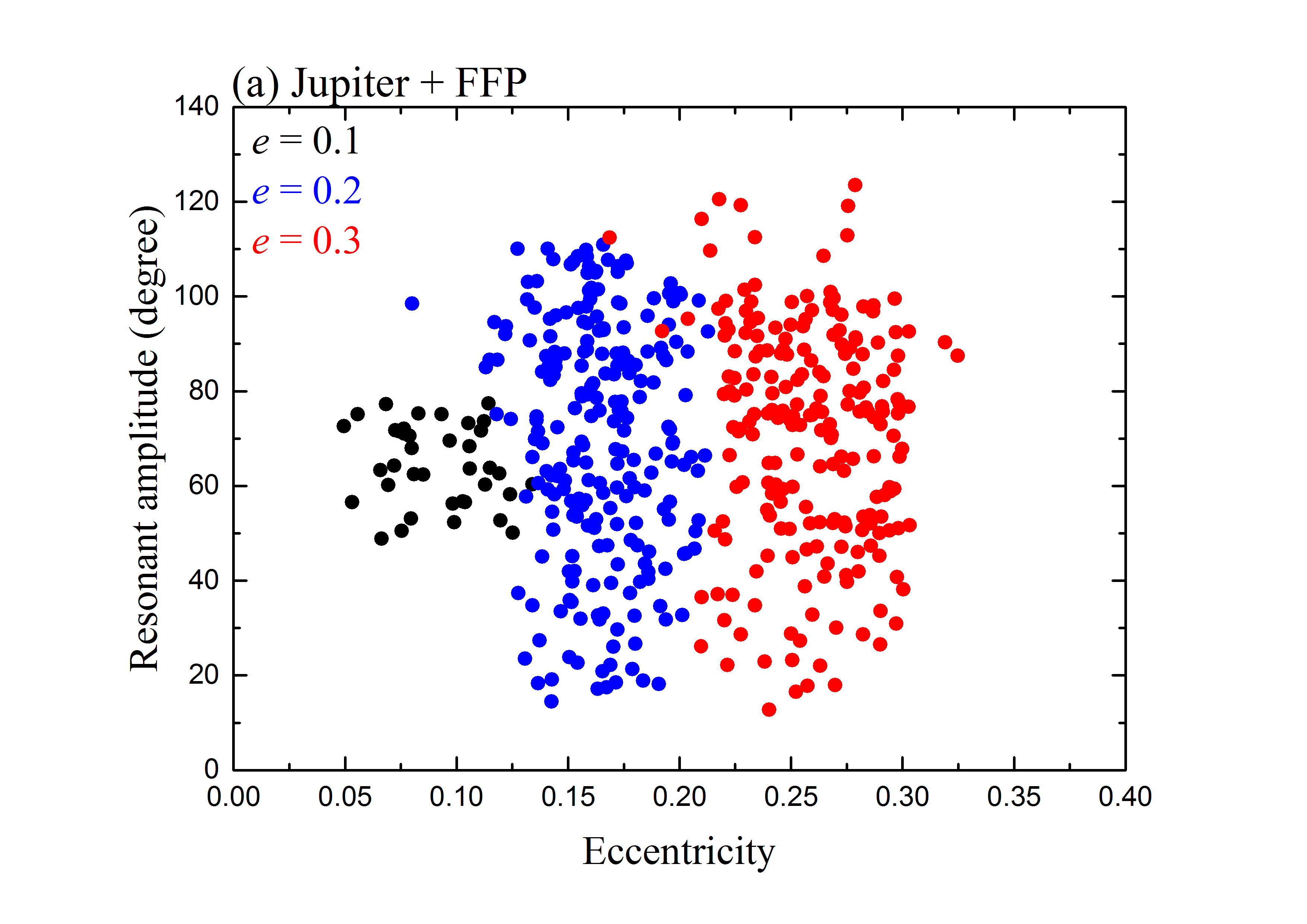}
  \includegraphics[width=9cm]{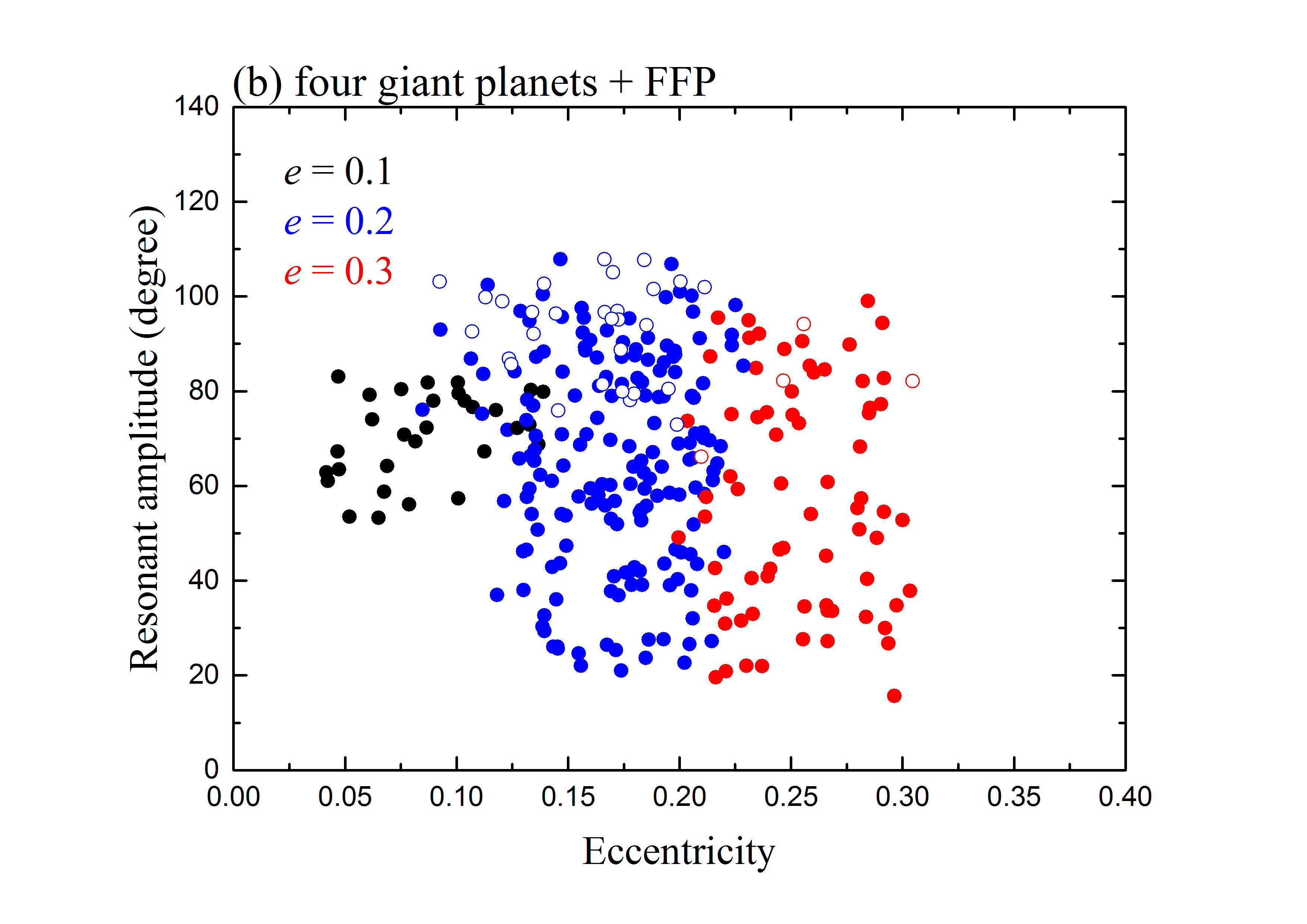}
  \end{minipage}
  \begin{minipage}[c]{1\textwidth}
  \vspace{0 cm}
  \includegraphics[width=9cm]{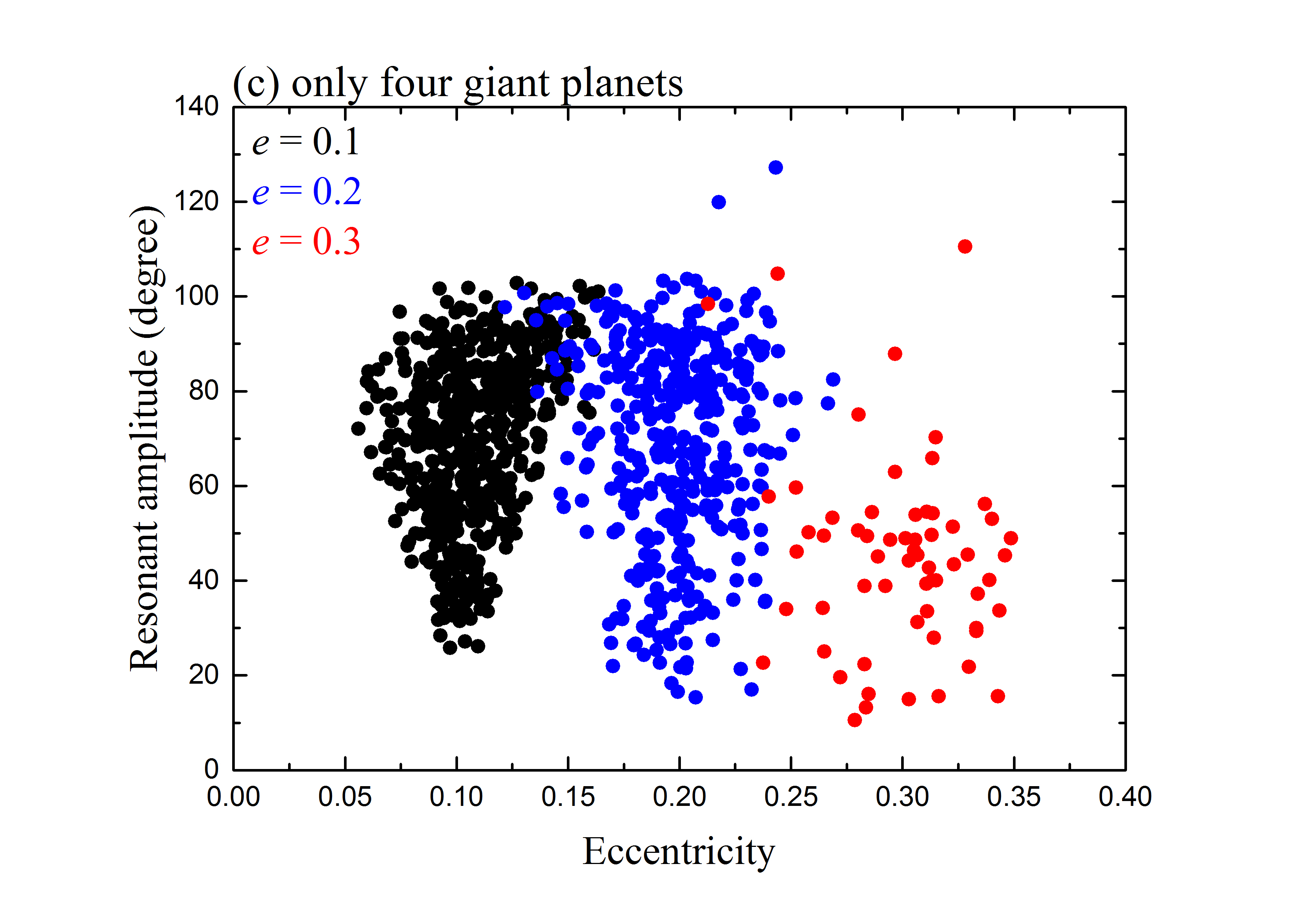}
  \includegraphics[width=9cm]{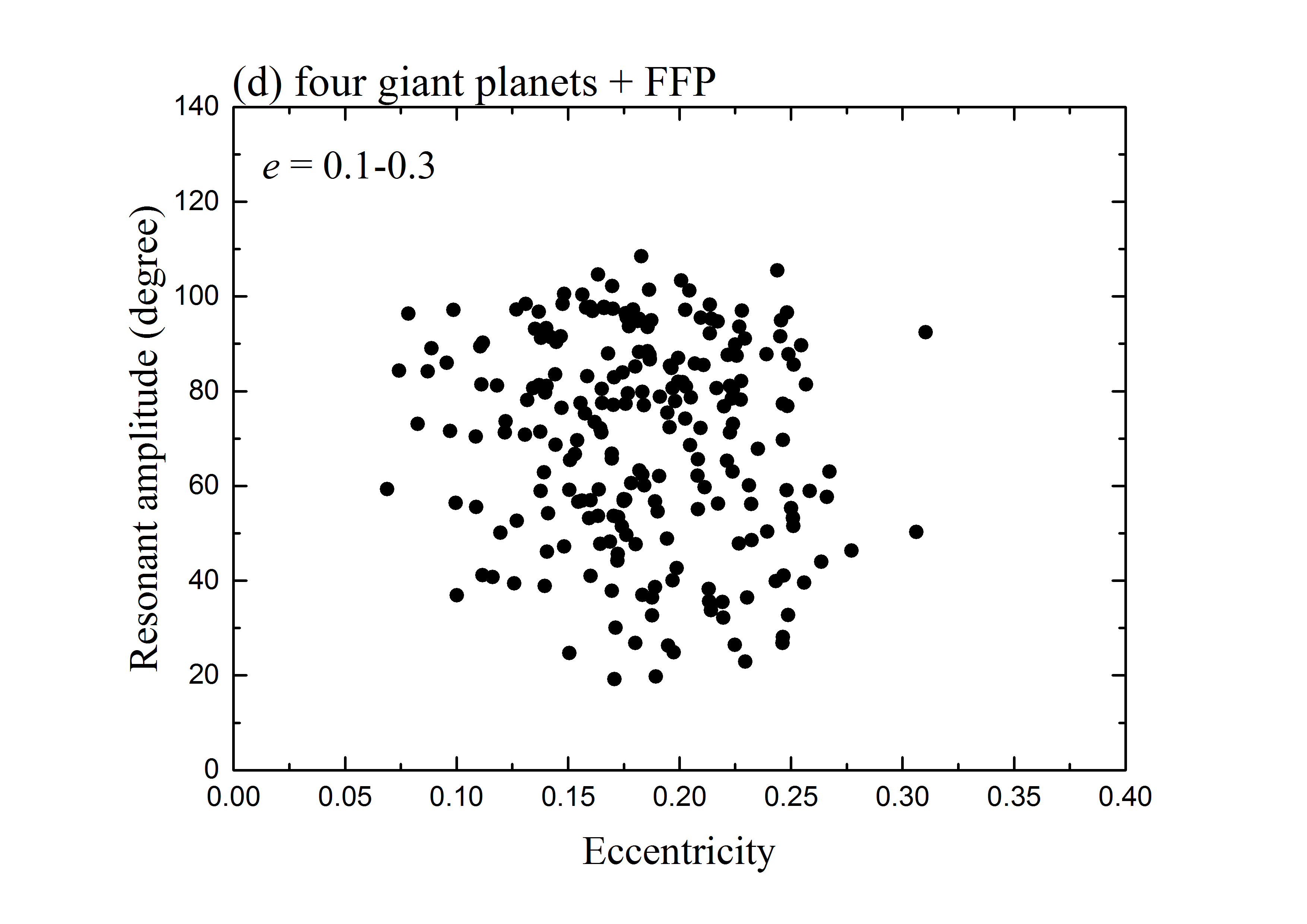}
  \end{minipage}
   \vspace{0 cm}
%\caption{Distribution of eccentricities and resonant amplitudes for the simulated Hildas at the end of 100 Myr evolution in different planetary models: (a) Jupiter with FFP, as in Fig. \ref{all}(b); (b) four giant planets with FFP, and for reference, the filled and open circles indicate objects having initial resonant amplitudes $>60^{\circ}$ and $<60^{\circ}$, respectively; (c) just four giant planets. The test Hildas have their initial semi-major axes and resonant angles randomly distributed across the entire phase space of the 3:2 Jovian resonance. In panel (b), subjected to the FFP flyby and the perturbations of the four giant planets, the resonant amplitude distribution of simulated Hildas best resembles the observed one, as shown in Fig. \ref{all}(a).}
\caption{Distribution of eccentricities and resonant amplitudes for the simulated Hildas at the end of the 100 Myr evolution in different planetary models: (a) Jupiter with FFP, as in Fig. \ref{all} (b); (b) four giant planets with FFP, and for reference, the filled and open circles indicate objects having initial resonant amplitudes $>60^{\circ}$ and $<60^{\circ}$, respectively; (c) just four giant planets. The test Hildas have their initial semi-major axes and resonant angles randomly distributed across the entire phase space of the 3:2 Jovian resonance; (d) similar to panel (b), but the test Hildas have initial eccentricities varying randomly within the interval $0.1<e<0.3$. In panels (b) and (d), subjected to the FFP flyby and the perturbations of the four giant planets, the resonant amplitude distribution of simulated Hildas best resembles the observed one, as shown in Fig. \ref{all} (a).}
  \label{randomEcc}
\end{figure*}

\begin{figure*}
  \centering
  \begin{minipage}[c]{1\textwidth}
  \vspace{0 cm}
  \includegraphics[width=8cm]{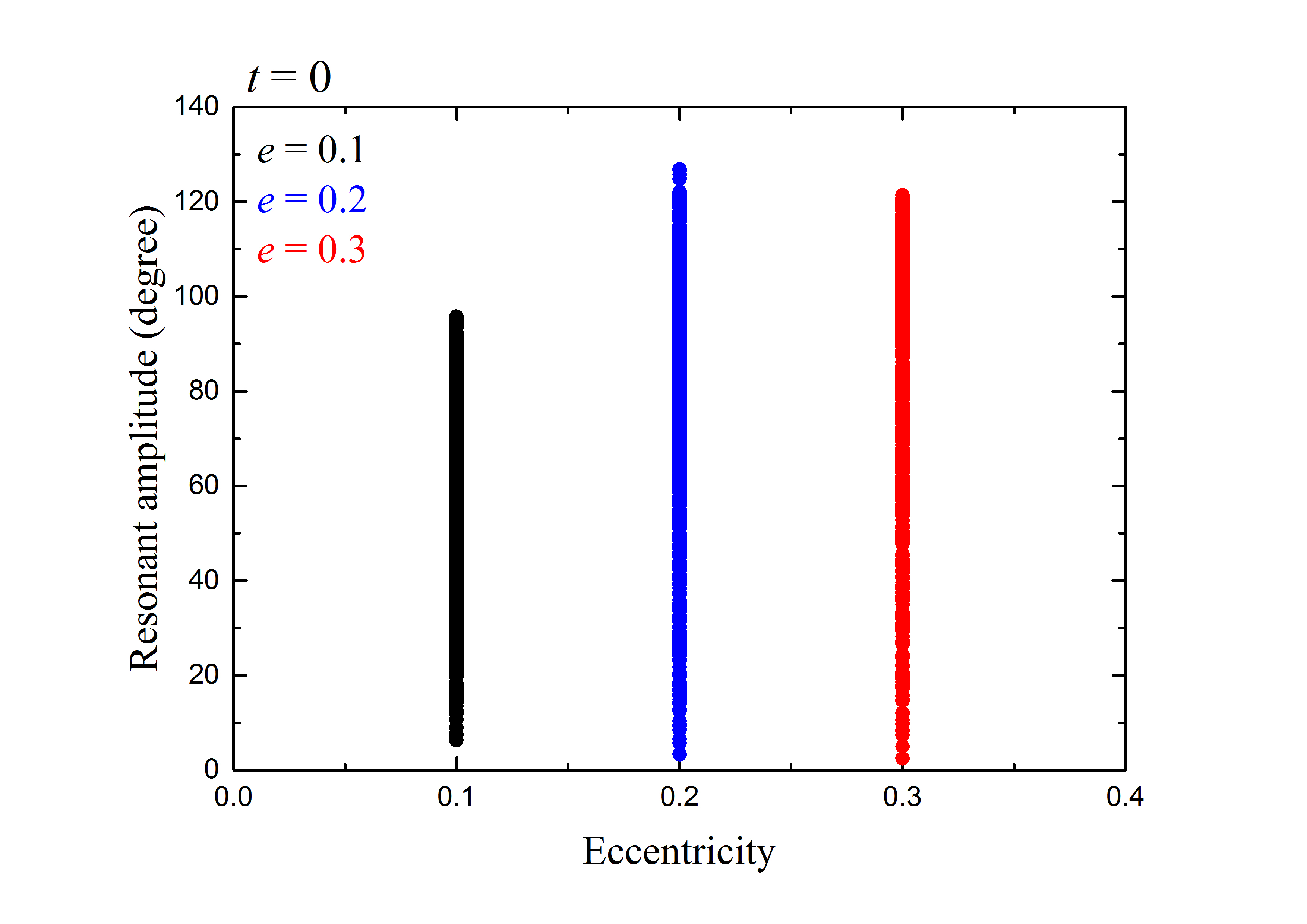}
  \includegraphics[width=8cm]{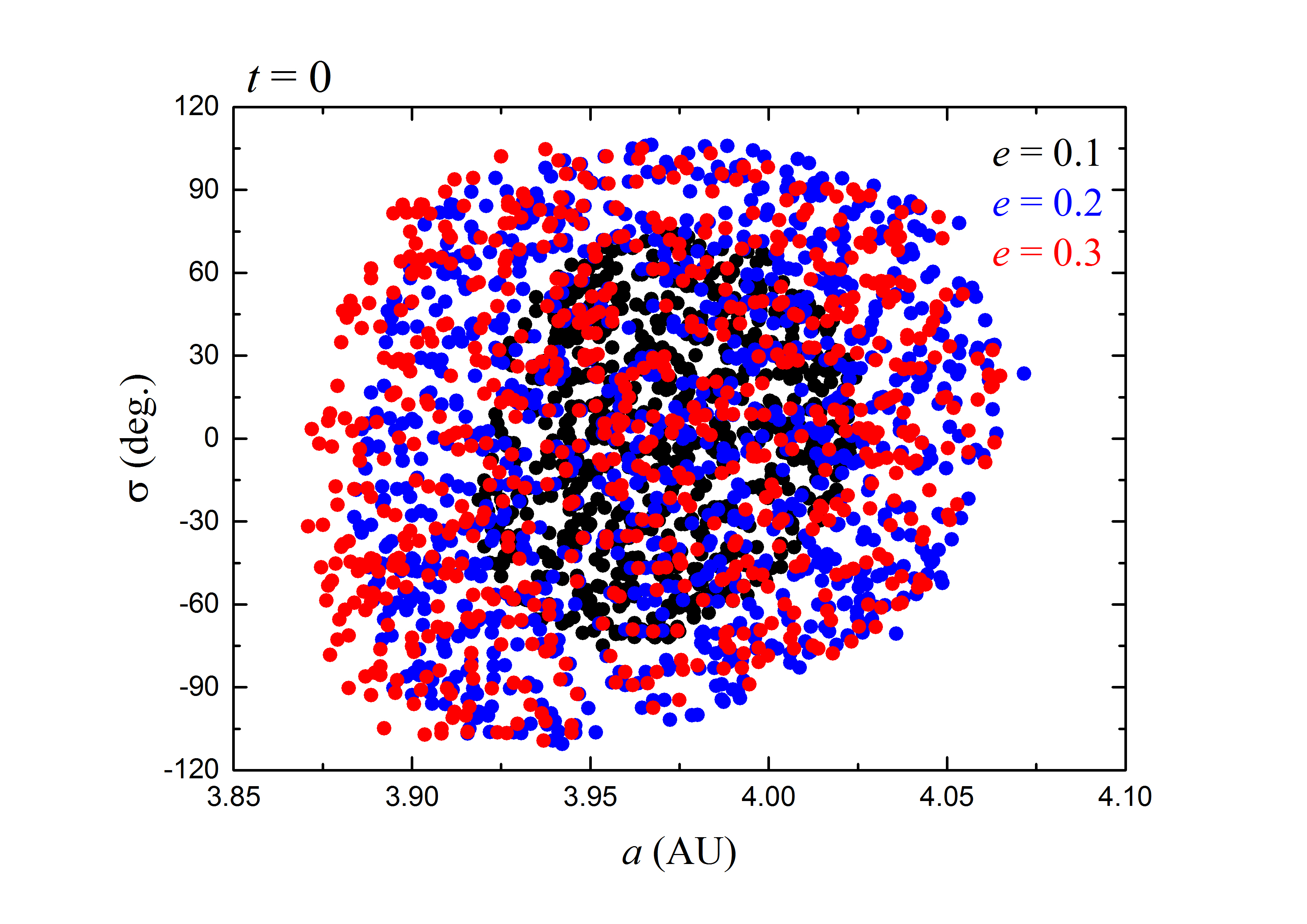}
  \end{minipage}
  \begin{minipage}[c]{1\textwidth}
  \vspace{-0.3 cm}
  \includegraphics[width=8cm]{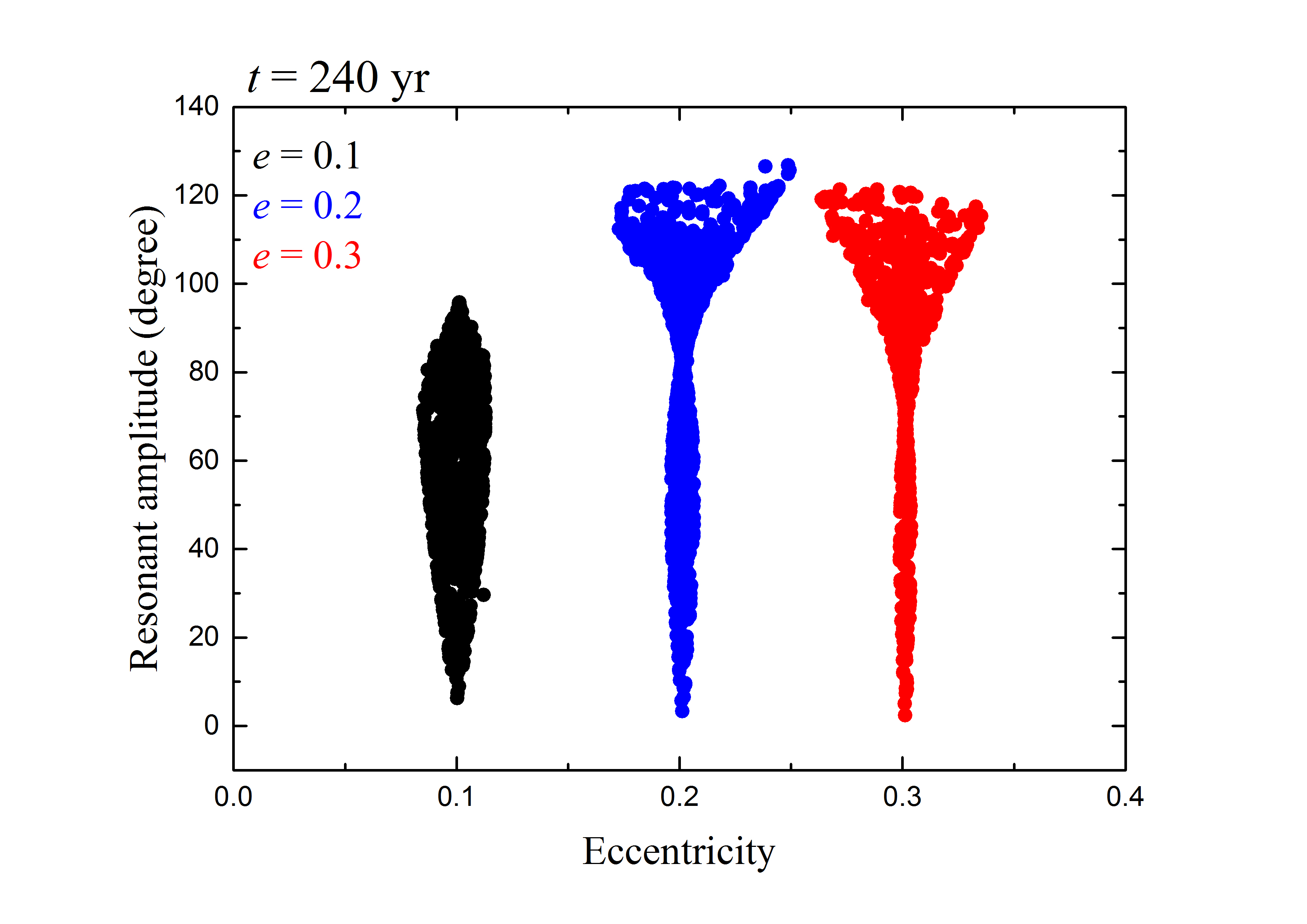}
  \includegraphics[width=8cm]{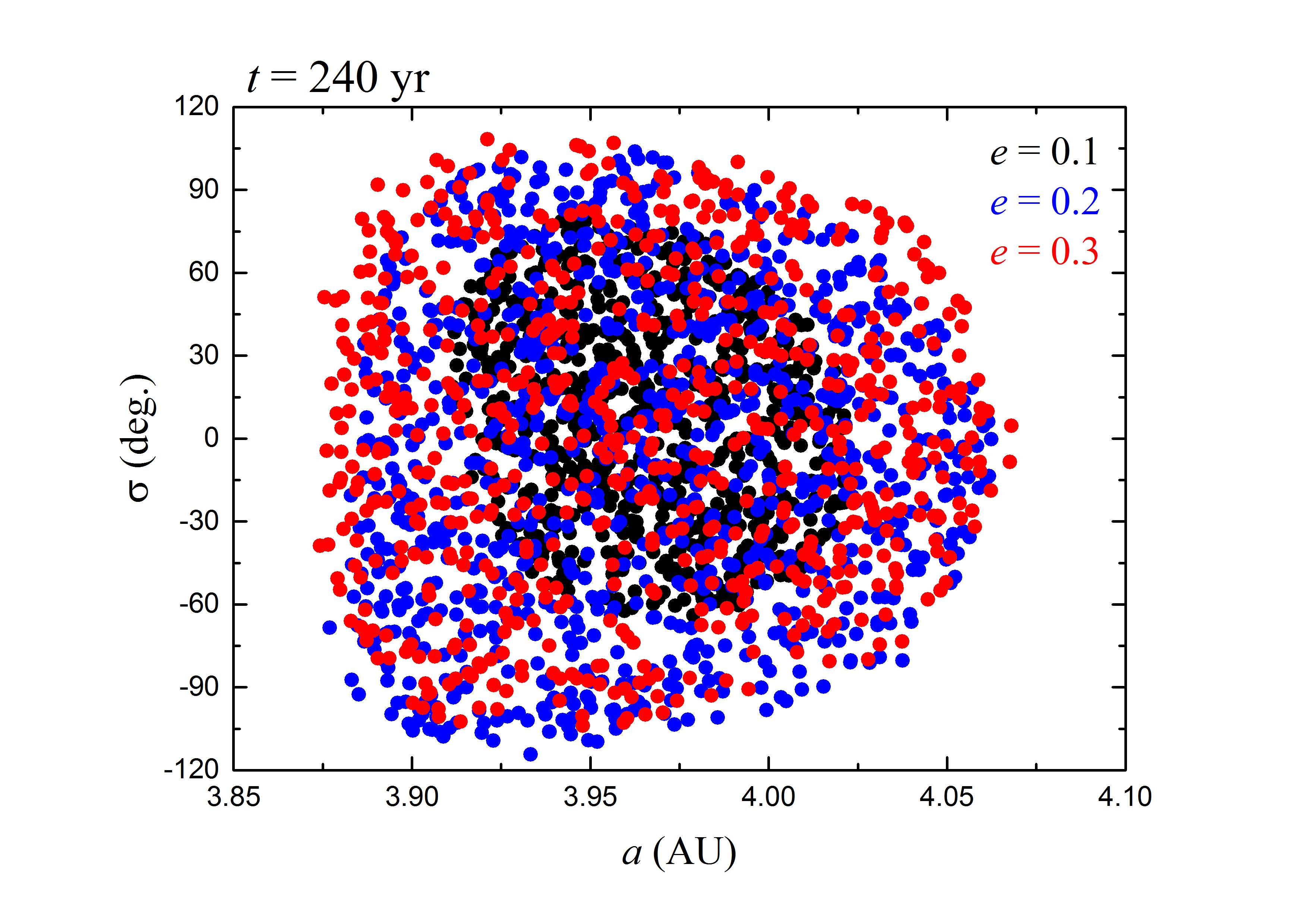}
  \end{minipage}
  \begin{minipage}[c]{1\textwidth}
  \vspace{-0.3 cm}
  \includegraphics[width=8cm]{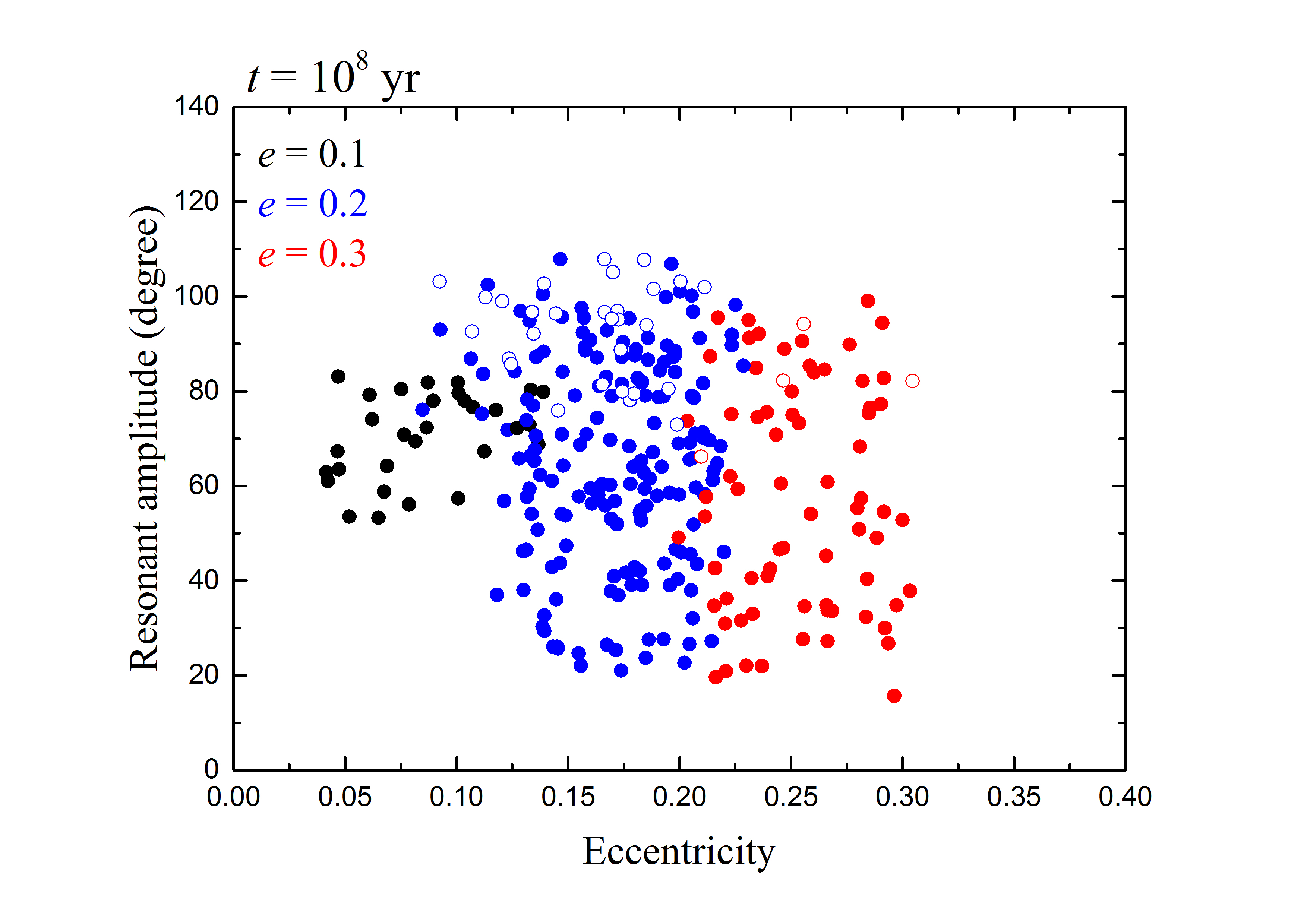}
  \includegraphics[width=8cm]{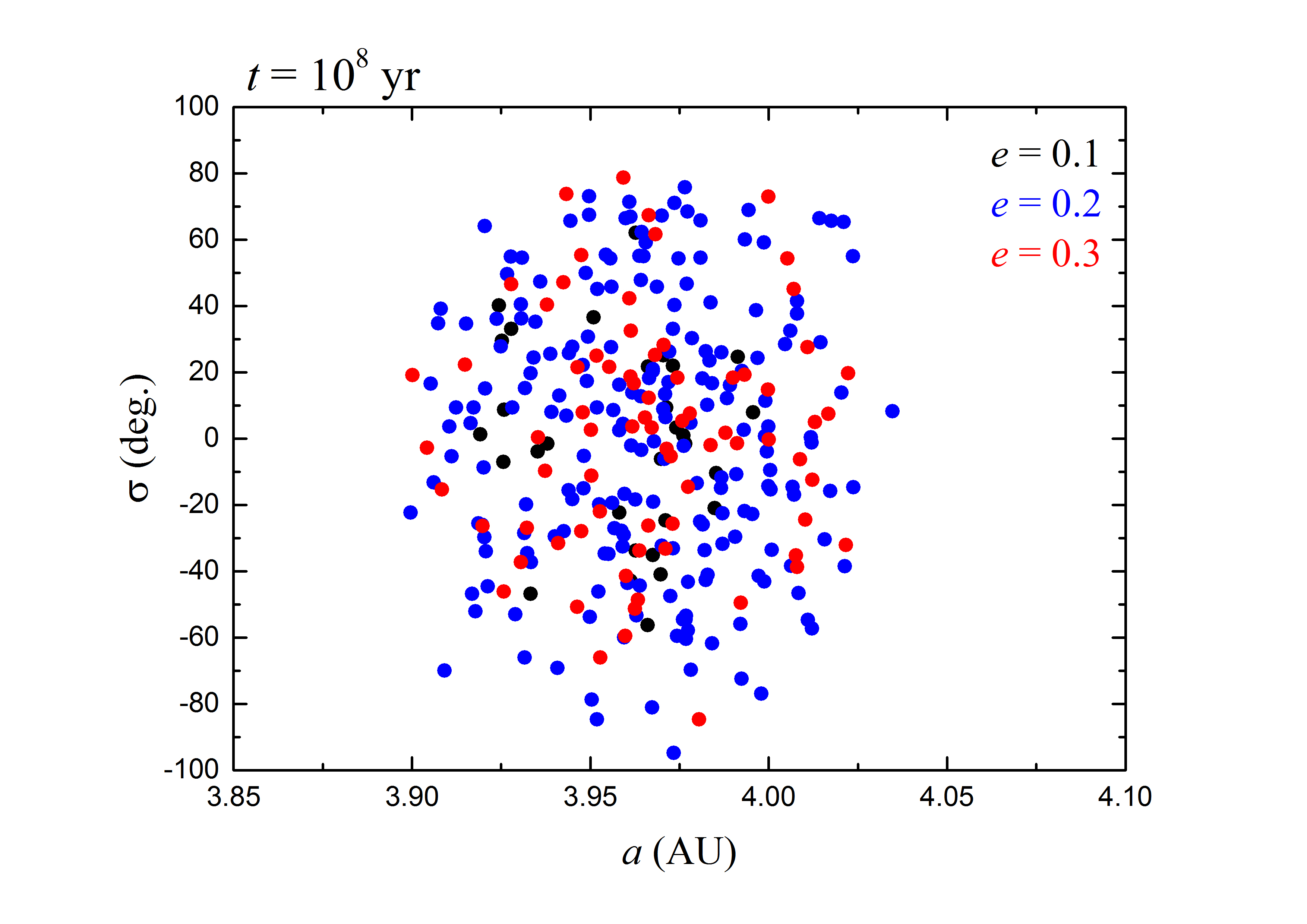}
  \end{minipage}
   \vspace{0 cm}
\caption{Snapshots of eccentricity vs. resonant amplitude (left column) and semi-major axis $a$ vs. resonant angle $\sigma$ (right column), taken from our simulations with fixed $e=0.1$, 0.2, and 0.3 in the model consisting of the FFP and four giant planets (i.e. the case considered in Fig. \ref{randomEcc} (b)). Each row represents the distribution of the simulated Hildas at three different time epochs: (top panels) at the beginning of the simulations ($t=0$); (middle panels) after the initial relaxation phase but just before the FFP event ($t=240$ yr); (bottom panels) at the end of the 100 Myr evolution. Note that the bottom left panel is exactly the same as Fig. \ref{randomEcc} (b).}
  \label{t0t240tf}
\end{figure*}

%Snapshots of eccentricity vs. resonant amplitude (left column) and semi-major axis $a$ vs. resonant angle $\sigma$, taken from our simulation with fixed $e=0.1$, 0.2, and 0.3 in the model of four giant planets with FFP (i.e. the case considered for \ref{randomEcc}(b)). Each row represents the distributions of the simulated Hildas at three different epochs: (top panels) the beginning of simulation ($t=0$); (middle panels), after the initial relaxation phase but just before the FFP flyby ($t=240$ yr); c, the end of 100 Myr evolution. Note that the bottom left panel is exactly the same figure as Fig. \ref{randomEcc}(b).

To firmly verify our replication of the resonant amplitude distribution of Hilda asteroids, we extend the numerical simulations performed above with two improvements. The first one is that the simulations now include all test Hildas over the entire $a$-$\sigma$ phase space of the 3:2 Jovian resonance, rather than just the specific samples from Populations \Rmnum1 and \Rmnum2. 
%Based on the theoretical resonant structure presented in Fig. \ref{width}, we select the initial semi-major axes $a$ of test Hildas randomly in the range of $[a_{in}, a_{out}]$, where $a_{in}$ and $a_{out}$ denote the inner and outer boundaries at a given $e$, respectively (see the black curves). 
Based on the theoretical analysis of the resonant structure presented in Sect. 3 below, we randomly select the initial semi-major axes $a$ of test Hildas within the range $[a_{in}, a_{out}]$, where $a_{in}$ and $a_{out}$ denote the inner and outer boundaries in $a$ of the 3:2 Jovian resonance.
The initial resonant angles $\sigma$ are randomly chosen from $-150^{\circ}$ to $150^{\circ}$. In each case of $e=0.1$, 0.2, and 0.3, we introduce 3000 test Hildas with given $a$ and $\sigma$ for a run. 

As we did before, these test Hildas are first confirmed to be librating objects through the pre-runs that do not include the FFP. Subsequently, they are regarded as the initial samples in the FFP flyby model for the 100 Myr long-term orbital calculations. The final distribution of eccentricities and resonant amplitudes for the simulated Hildas is shown in Fig. \ref{randomEcc} (a). In comparison with Fig. \ref{all} (a) for the observed Hildas, we observe a similar pattern, i.e. none of the objects with $e\lesssim 0.1$ have $A<40^{\circ}$, but there is a slight discrepancy that a small portion of them has $A<20^{\circ}$ at $e > 0.1$.

Similar to the numerical simulation carried out above, where only Jupiter is considered, we conduct the second improvement by incorporating perturbations from the other giant planets (Saturn, Uranus, and Neptune), also for a 100 Myr evolution. Under the perturbations of all four giant planets, we performed two more simulations, with and without the FFP flyby, and the corresponding distributions of simulated Hildas are shown in Figs. \ref{randomEcc} (b) and (c), respectively. We see in panel (b) that the Hilda distribution associated with the FFP flyby aligns more closely with observations, surpassing even the match shown in panel (a), as nearly all objects have $A>20^{\circ}$ regardless of $e$.

In fact, the difference between Figs. \ref{randomEcc} (a) and (b) clearly shows that there is a notable effect induced by Saturn, Uranus, and Neptune. As shown in Fig. \ref{randomEcc} (c), in the case of no FFP, the secular perturbations of these three giant planets do excite the resonant amplitudes of Hildas, reaching values of $A\gtrsim26^{\circ}$ at $e=0.1$ and $A\gtrsim12^{\circ}$ at $e=0.2$-0.3. However, such excitation does not appear large enough to replicate the observed patterns. This indicates that planetary perturbations in the current structure of the Solar System may not alone account for the $A$-distribution of the Hildas.

So far, the eccentricities of the test Hildas have been fixed at three representative values: $e=0.1$, 0.2, and 0.3. As a supplement, a new experiment has been conducted, letting $e$ in the initial conditions to vary randomly within the interval $(0.1, 0.3)$, in the framework that includes the FFP and four giant planets. As shown in Fig. \ref{randomEcc} (d), the distribution of eccentricities and resonant amplitudes for the simulated Hildas at the end of the 100 Myr evolution is nearly the same as that displayed in Fig. \ref{randomEcc} (b). Nevertheless, one may notice a slight discrepancy: around $e=0.1$ and $e=0.3$, the number density of the simulated Hildas appears lower. This is easy to understand because the test Hildas start from a uniform eccentricity distribution within the interval $0.1<e<0.3$. Naturally, the number of test Hildas with initial $e\sim0.1$ and $e\sim0.3$ is much smaller compared to the number used for Fig. \ref{randomEcc} (b). In fact, combining the fixed-$e$ case and the random-$e$ case by superposing Fig. \ref{randomEcc} (d) onto Fig. \ref{randomEcc} (b) would result in a portrait more similar to Fig. \ref{all} (a) for the observed Hildas.

To clearly illustrate what happens during the FFP flyby event--not just how the test Hildas are distributed in the eccentricity and resonance amplitude plane after the FFP crossing--we provide additional plots at three time epoch: (1) at time $t=0$, (2) at time $t=240$ yr, i.e. after the initial relaxation phase but before the FFP event (as described in Sect. 2.3), and (3) the final distributions. For the sake of clarity, we focus on the experiment with fixed $e=0.1$, 0.2, and 0.3 in the model of four giant planets with FFP, as this experiment is most representative and highlights differences at various $e$ values. The plots are shown in Fig. \ref{t0t240tf}, where the bottom left panel is exactly the same figure as Fig. \ref{randomEcc} (b). At each time epoch, in addition to the distribution of eccentricity and resonance amplitude shown in the left column, we also present the distribution of particles in the $(a, \sigma)$ plane in the right column. It is important to note that the $a$-ranges of the test Hildas with $e=0.2$ (blue dots) and $e=0.3$ (red dots) are much wider than those with $e=0.1$ (black dots). This pattern helps correlate the obtained numerical results with the theoretical phase portrait of the 3:2 Jovian MMR that we will show below in Sect. 4. We would like to mention that, even with a longer relaxation time before the FFP event, the $(a, \sigma)$ distribution remains nearly identical. This is because the test Hildas are chosen to cover the entire $(a, \sigma)$ space of the 3:2 Jovian MMR. Although $e$ tends to be slightly more dispersed than what is shown in the middle left panel of Fig. \ref{t0t240tf}, the final distribution of Hildas can hardly be affected, as the $e$-relaxation can be effectively represented by the random-$e$ case we have considered. This suggests that the duration of the relaxation time is largely irrelevant.

In our calculations spanning 100 Myr, we find that the Hildas would achieve stable resonant configurations within 10 Myr of evolution, as their resonant amplitudes are unlikely to change significantly thereafter in the context of just four giant planets. However, it is possible that before the arrival of the FFP, the initial $A$-distribution of the Hildas from formation models could differ from our current assumptions, potentially not occupying the entire $A$ range. We will discuss this consideration in more detail in the following subsection.

\subsection{Dependence on initial $A$-distribution}

%In the study of this paper, we assume that the test Hildas have initial resonant amplitudes covering the entire range of $A\le150^{\circ}$. This assumption, denoted as Case 1, is reasonable if the Hildas were formed in situ. While the formation of Hildas may undergo complex dynamical processes, resulting in $A$-distributions of test Hildas that vary across different formation models.

%Under the classical planet migration scenario of \citet{Malh1995}, in which Jupiter's orbit experiences smooth radial shift, \citet{Fran04} studied the capture of field asteroids into Hilda-type orbits in the 3:2 Jovian MMR. Their outcomes show that the a small fraction of captured Hildas were populated in the resonant amplitude region of $A<20^{\circ}$. For such less excited Hildas, denoted as Case 2, they would escape the 3:2 Jovian MMR in the context of our FFP flyby hypothesis, just as what we found in Case 1. Therefore, the absence of Hildas with $A$ below 20$^{\circ}$-40$^{\circ}$ will be reproduced. 

To date, the formation of Hildas remains uncertain, whether it occurred in situ or involved complex dynamical processes. Therefore, the initial $A$-distributions of test Hildas that we considered in the FFP flyby scenario could vary across different formation models. Some potential cases regarding the dependence of the resonant patterns of Hildas on their initial $A$-distributions are the following:

%The details of the dependence of the resonant patterns of Hildas on their initial $A$-distributions can be briefly as follows:

\underline{Case 1:} In this study, we assume that the test Hildas have initial resonant amplitudes covering the entire range of $0<A\le150^{\circ}$. This assumption is reasonable if the Hildas were formed in situ. As demonstrated earlier, Hildas with low resonant amplitudes would escape from the 3:2 Jovian MMR in the context of our FFP flyby hypothesis.

\underline{Case 2:} Under the classical planet migration scenario of \citet{Malh1995}, where Jupiter's orbit undergoes a smooth radial shift, \citet{Fran04} studied the capture of field asteroids into Hilda-type orbits in the 3:2 Jovian MMR. Their findings indicate that a small fraction of captured Hildas occupied the resonant amplitude region of $A<20^{\circ}$. Consequently, these less excited Hildas would disappear in the FFP flyby scenario, similar to what was stated in Case 1. Therefore, the absence of Hildas with $A$ below $20^{\circ}$-$40^{\circ}$ will be replicated. 

\underline{Case 3:} Previous studies, such as \citet{Broz08} and \citet{Roig15}, have shown that a catastrophic instability of Hilda asteroids would occur when Jupiter and Saturn crossed their mutual 1:2 or 2:3 MMR. As a result, primordial Hildas were effectively removed from their orbits. After this instability, some main belt asteroids became implanted in the Jovian 3:2 resonance, forming the Hildas observed today. We hypothesize that this violent removal and implantation process increased the resonant amplitudes of the Hildas. It is very interesting to point out that, even if the Hildas were initially overexcited, for instance, settling into resonant orbits with medium to large $A>60^{\circ}$, our FFP flyby scenario is still capable of reshaping such an $A$-distribution to align with observations. The underlying mechanism is the jump of the Jovian 3:2 resonance, which is coupled with the Jupiter's migration caused by the FFP flyby. 
%The removal/implantation origin of Hildas will result in a distinct resonant amplitude distribution, different than if the Hildas were captured by a smoothly migrating Jupiter mentioned in Case 2. Considering the violent 
%For detailed studies of this removal/implantation origin of the Hildas, the reader is referred to the two articles \citet{Broz08} and \citet{Roig15}.

As illustrated in Sect. \ref{mechanism}, the Hildas initially with $A>60^{\circ}$ considered in this Case 3 would end up with relatively small $A$. In particular, Hildas starting with $A\sim90^{\circ}$ could evolve to orbits with $A\sim20^{\circ}$-$40^{\circ}$. To visualize this redistribution of resonant amplitudes, in Fig. \ref{randomEcc} (b), we marked the objects with initial $A>60^{\circ}$ using filled circles. It can be seen that after the FFP flyby and the 3:2 resonance jump, the vast majority of the Hilda population consists of these objects. From this, we infer that despite the uncertain formation history, the resonant amplitude region from medium to large $A$ should consistently be well populated with Hildas. Thus, this subset of the Hilda population alone could reproduce the overall $A$-distribution similar to the distinct observational patterns examined in this work.

\underline{Case 4:} If the formation process can precisely reshape the Hildas to match observational resonant patterns, specifically the absence of objects with $A<20^{\circ}$-$40^{\circ}$, our FFP flyby scenario would not lead to an $A$-distribution contradictory to observations. Similar to our analysis in Case 3 above, incorporating the FFP flyby mechanism shows that the surviving Hildas after the jump of the 3:2 Jovian MMR are primarily composed of those with initial $A>60^{\circ}$, and their final $A$-distribution resembles observations. On the other hand, objects from the smaller initial $A$ region only contribute somewhat to the Hildas with large $A$ values ($80^{\circ}$-$100^{\circ}$), as indicated by the open circles in Fig. \ref{randomEcc} (b). Whether these objects are included or not, the overall $A$-distribution will not change. This suggests that even if the desert observed in the small $A$ region appeared during the formation period, it can still be preserved. In this case, while the FFP flyby may not be essential regarding Hildas, its use in explaining the number asymmetry of Jupiter Trojans remains unaffected.

The four cases discussed above provide more conclusive support for the FFP flyby scenario that we proposed to explain the number asymmetry of Jupiter Trojans. For Hildas originating from various formation models, whether their resonant amplitudes were less excited (Case 1 and Case 2), overexcited (Case 3), or excited to the right degree (Case 4), the FFP flyby scenario can always result in the absence of Hildas with $A<40^{\circ}$ at $e<0.1$ and in almost no objects having $A<20^{\circ}$ elsewhere. In other words, to achieve these resonant patterns consistent with observations, our FFP flyby mechanism is nearly independent of the initial range of resonant amplitudes of the Hildas. As long as a population with $A>60^{\circ}$ exists, the FFP flyby can always produce the $A$-distribution that does not contradict observations.

\section{Theoretical analysis}

The obtained numerical results suggest that the fundamental factor controlling the resonant amplitude distribution of the simulated Hildas is the resonant width, which depends on the eccentricity. In this section, a multi-harmonic Hamiltonian model is constructed for the 3:2 Jovian resonance. It allows us to study analytically the phase space structures at different eccentricities up to 0.3. We refer the reader to \citet{Lei20} for more details on the following calculations.

We consider the planar circular restricted three-body problem (CRTBP) that consists of the Sun, Jupiter, and a massless Hilda. In the context of this CRTBP, according to \citet{Lei20}, the multi-harmonic resonant Hamiltonian can be written as
\begin{equation}\label{Eq1}
{{\cal H}} =  - \frac{{{\mu ^2}}}{{2{{\left( {k{\Phi _1}} \right)}^2}}} - {k_J}{n_J}{\Phi _1} - \sum\limits_{n = 0}^N {{{\cal C}_n}\left( {{\Phi _1},{\Phi _2}} \right)\cos (n{\phi _1})},
\end{equation}
where $\mu$ is the gravitational constant of the Sun, $n_J$ is the mean motion of Jupiter, ${\phi _1} = k\lambda  - {k_J}{\lambda _J} - ({k_J} - k)\varpi$ is the argument associated with the $k_J : k$ eccentricity-type resonance, ${\Phi _1} = \frac{1}{k}\sqrt{\mu a}$ is the conjugate momentum of ${\phi _1}$, ${\Phi _2} = \sqrt{\mu a} (\frac{{{k_J}}}{k}-\sqrt{1-e^2})$ is the motion integral, and ${\cal C}_n$ are certain coefficients defined in \citet{Lei20}. For the 3:2 Jovian interior resonance considered in this study, it holds that $k_J=3$ and $k=2$ and thus ${\phi _1}$ turns out to be resonant angle $\sigma$.

\begin{figure}
 \hspace{0cm}
  \centering
  \includegraphics[width=9cm]{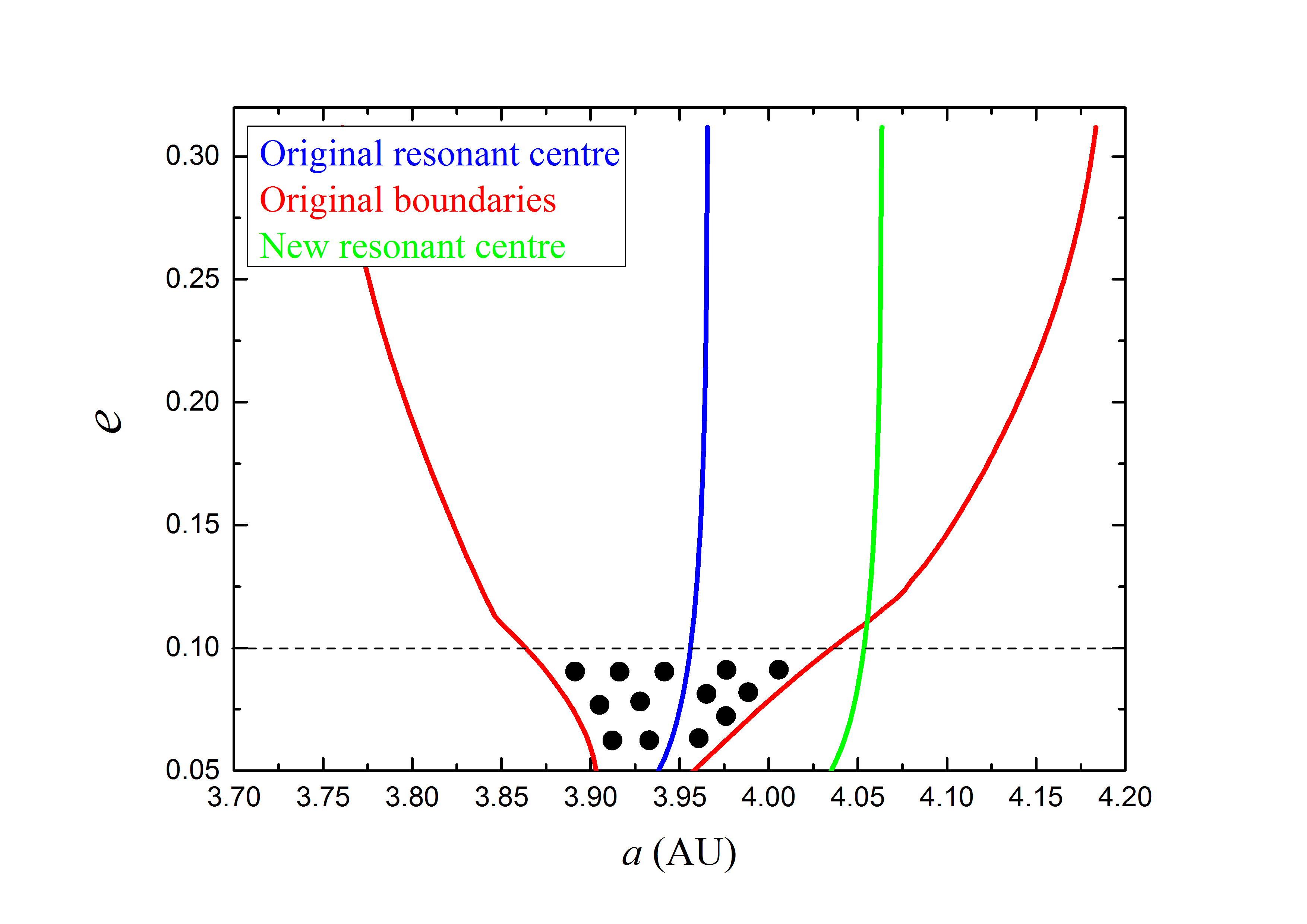}
  \caption{Resonant structure in the ($a$, $e$) plane for the 3:2 Jovian resonance. The red curves indicate the original boundaries of the stable libration zone and the blue curve indicates the original resonant centre, while the green curve represents the new resonant centre after the event of the FFP flyby. The initial test Hildas with $e\le0.1$ are shown as dots, which reside between the two red boundaries.}
  \label{width}
\end{figure}

When the motion integral ${\Phi _2}$ is given, the resonant Hamiltonian specified by Eq. (\ref{Eq1}) determines an integrable dynamical model. In particular, the equilibrium points of the resonant model are determined by
\begin{equation}\label{Eq2}
{\dot {\phi _1}} = \frac{{\partial {{\cal H}}}}{{\partial {\Phi _1}}},\quad {\dot \Phi _1} = -\frac{{\partial {{\cal H}}}}{{\partial {\phi _1}}},
\end{equation}
where the stable equilibrium points, denoted by $(\phi_{1,s},\Phi_{1,s})$, correspond to resonant centres, while the unstable equilibrium points, denoted by $(\phi_{1,u},\Phi_{1,u})$, correspond to saddle points. Level curves of resonant Hamiltonian passing through saddle points are referred to as dynamical separatrices (i.e. the resonant boundaries), which divide the phase space into regions of libration and circulation. Accordingly, the resonant boundaries can be evaluated from
\begin{equation}\label{Eq3}
{\cal H}(\phi_{1,u},\Phi_{1,u}) = {\cal H}(\phi_{1,s},\Phi_{1,\rm out}) = {\cal H}(\phi_{1,s},\Phi_{1,\rm in}),
\end{equation}
where $\Phi_{1,\rm in}$ and $\Phi_{1,\rm out}$ stand for the inner and outer boundaries in terms of the semi-major axis $a$, respectively. In addition, the location of the resonant centre $(\Phi_{1,s}, \Phi_2)$ can be equivalently expressed as $(a_c,e_c)$.

We then use our Hamiltonian model to analytically determine the resonant centre $(a_c,e_c)$ and resonant boundaries in terms of $a$ for the 3:2 Jovian resonance. As shown in Fig. \ref{width}, the blue and red curves indicate the original resonant centre and boundaries, respectively. It is obvious that the resonant width, i.e. the distance between the two boundaries, becomes narrower as the eccentricity decreases. After the strong perturbation caused by the FFP, Jupiter jumps outwards together with its 3:2 resonance, resulting in a new resonant centre (indicated by the green curve), which is beyond the original centre. This new resonant centre can be proportionately obtained by considering the expansion of Jupiter's orbit by $0.12$ AU.

In Fig. \ref{width} we specifically note that, at $e\le0.1$, the new resonant centre is exterior to the original libration zone enclosed by the two red boundaries where the test Hildas reside (indicated by the dots). As a result, after the onset of the FFP's strong perturbation and the jump of Jupiter, the test Hildas with initial $e=0.1$ are currently located at a distance from the new resonant centre and they would switch to the orbits with quite large $A$, as we found in the numerical simulations. In contrast, for the cases of $e=0.2$ and 0.3, the original libration zone encompasses the new resonant centre, around which the test Hildas could evolve to the orbits with relatively smaller $A$ due to a sudden change in the location of the 3:2 Jovian resonance.

\begin{figure}
  \centering
  \begin{minipage}[c]{0.5\textwidth}
  \centering
  \vspace{0cm}
  \includegraphics[width=8.2cm]{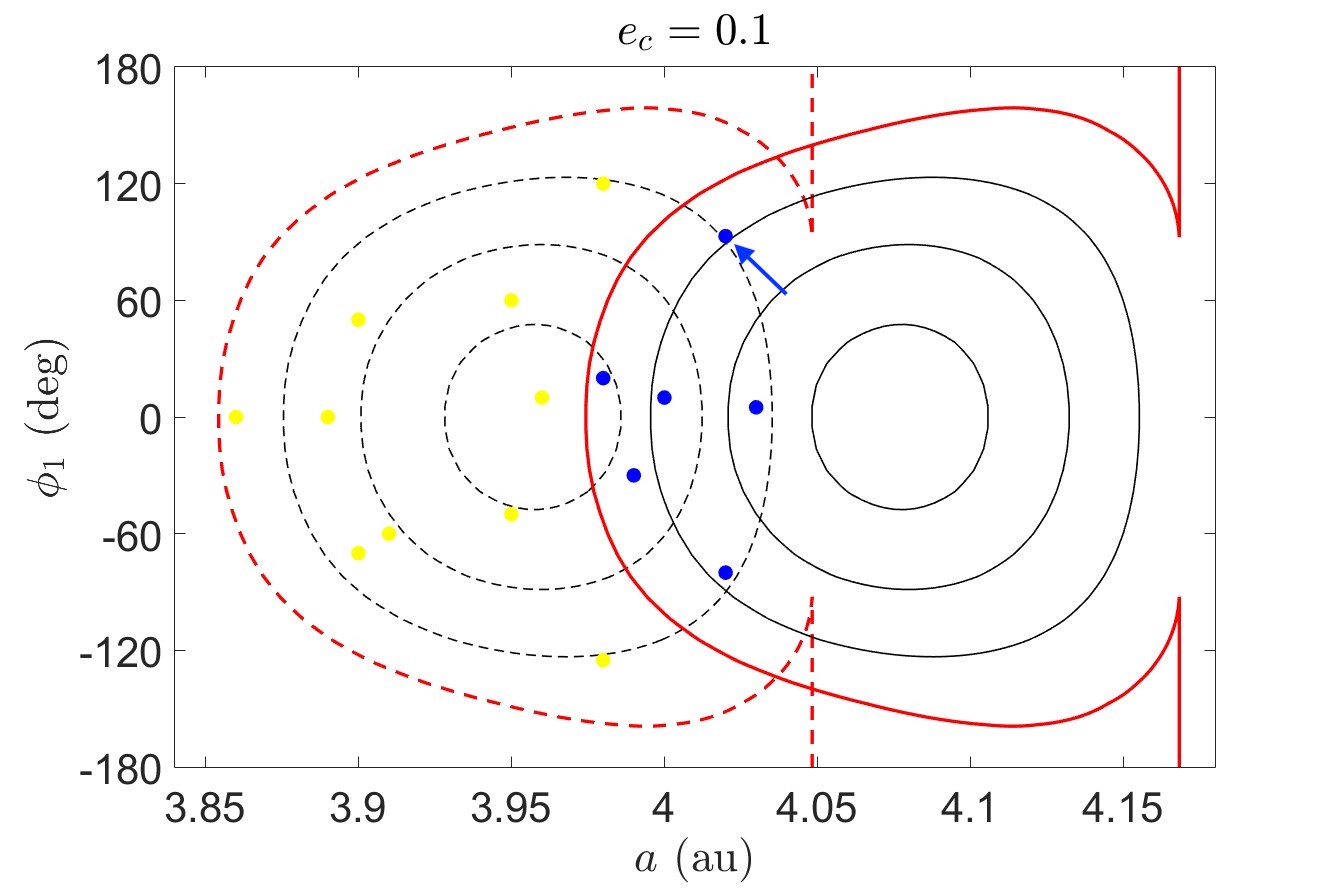}
  \end{minipage}
  \begin{minipage}[c]{0.5\textwidth}
  \centering
  \vspace{0.39cm}
  \includegraphics[width=8.2cm]{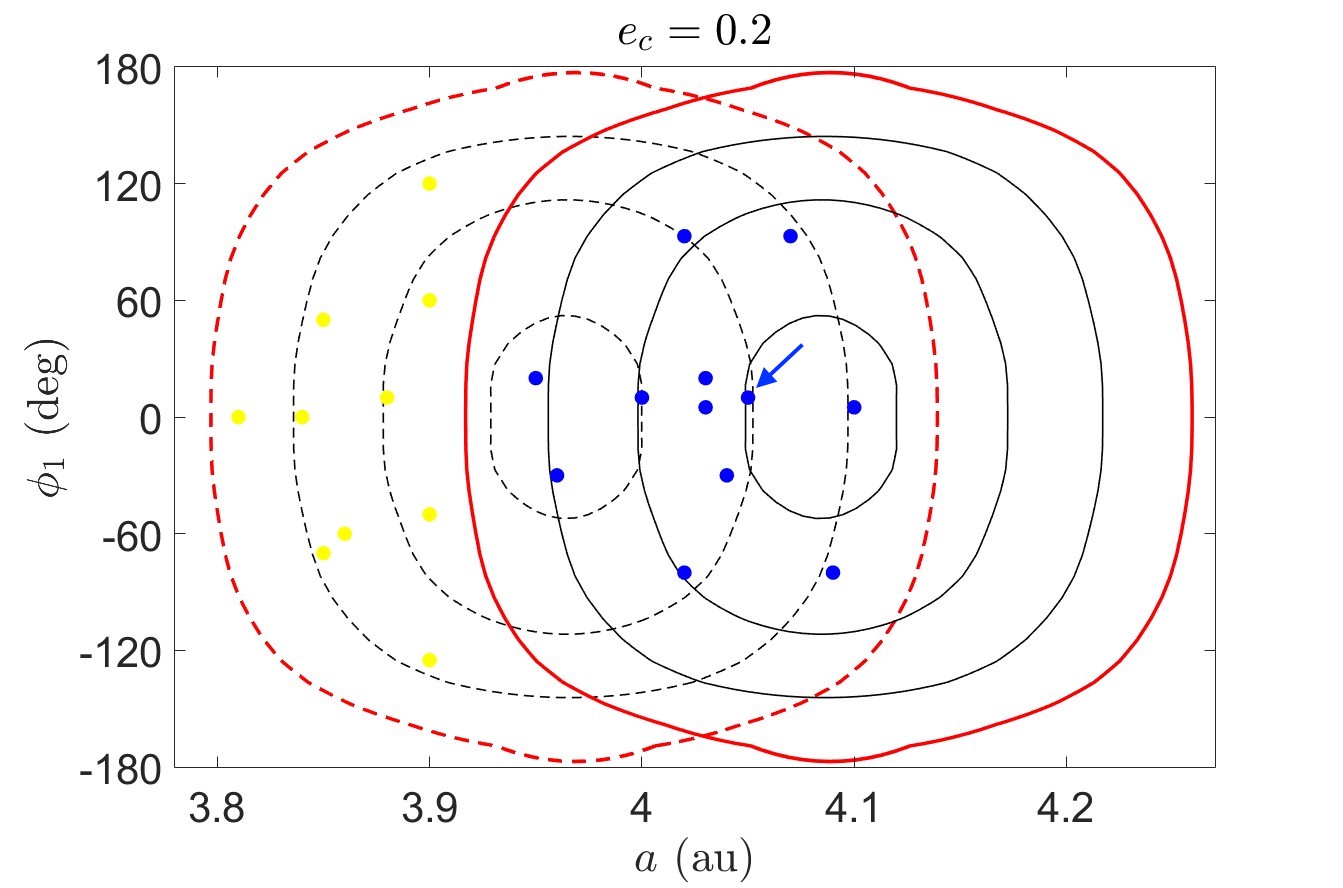}
  \end{minipage}
  \begin{minipage}[c]{0.5\textwidth}
  \centering
  \vspace{0.39cm}
  \includegraphics[width=8.2cm]{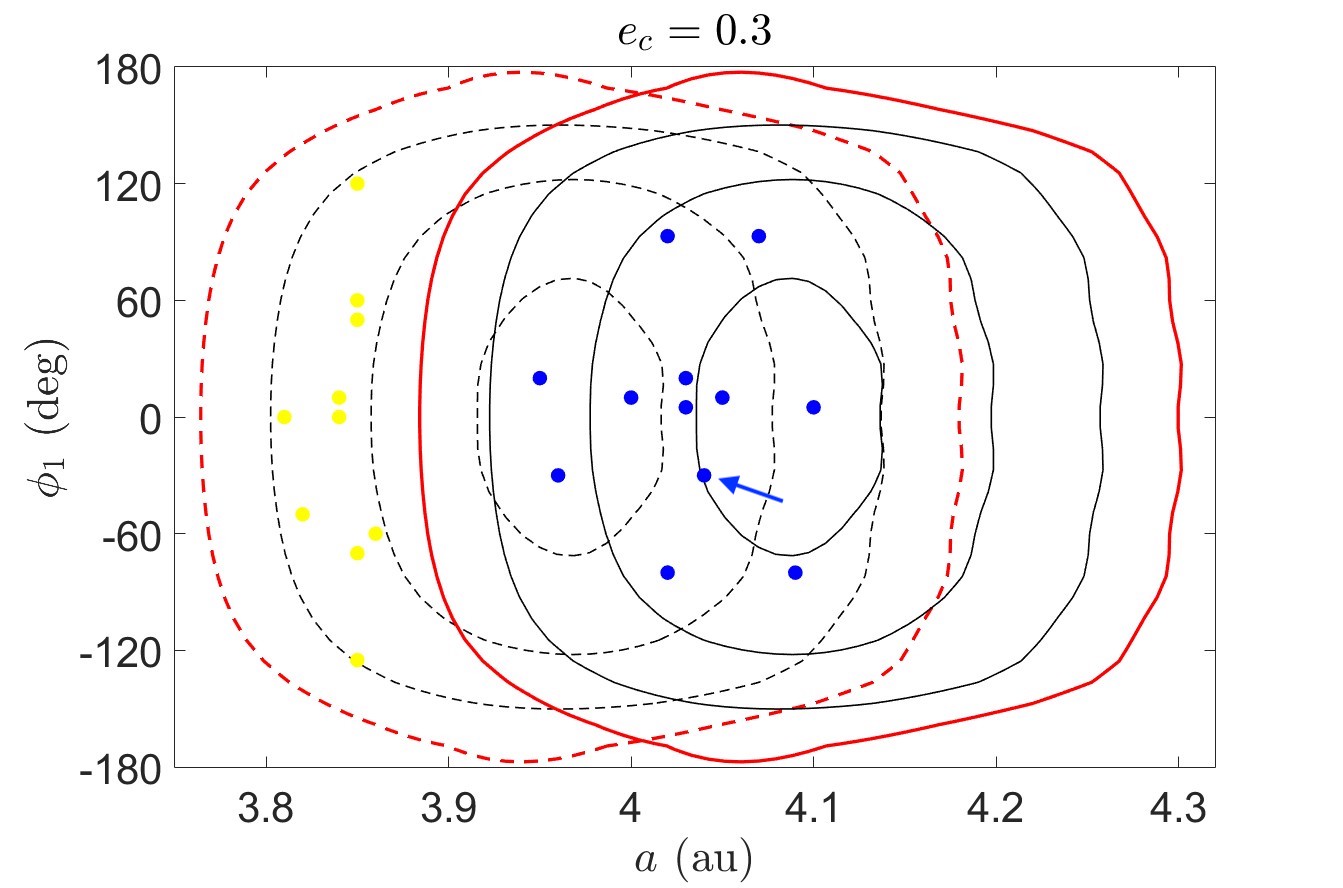}
  \end{minipage}
% \caption{Phase portraits of the 3:2 Jovian resonance in the $(a,\phi_1)$ plane for three considered eccentricities of $e_c=$0.1 (top panel), 0.2 (middle panel), and 0.3 (bottom panel). The black and red curves denote the resonant islands and the resonant boundaries, respectively. The dashed and solid curves refer to the original and new resonances as we noted in Fig. \ref{width}. The dots stand for the test Hildas.Note that the angle $\phi_1$ defined in Eq. \ref{Eq1} is actually the resonant angle $\sigma$ of the Hildas. }
\caption{Phase portraits of the 3:2 Jovian resonance in the $(a,\phi_1)$ plane for the three considered eccentricities of $e_c=0.1$ (top panel), 0.2 (middle panel), and 0.3 (bottom panel). The black and red curves denote the resonant islands and resonant boundaries, respectively. The dashed and solid curves refer to the original and new resonances, as noted in Fig. \ref{width}. The dots represent the test Hildas, which are used for illustration rather than as the actual simulation objects. Note that the angle $\phi_1$, defined in Eq. (\ref{Eq1}), is actually the resonant angle $\sigma$ of the Hildas.}
 \label{aphi1}
\end{figure}

%For better visualization of the `shift of resonant centre' scenario, we would show the actual phase portraits of the resonance in the $(a,\phi_1)$ plane. Note that this plane is equivalent to the `canonical' $(\Phi_1,\phi_1)$ plane, but more intuitive. The $(a,\phi_1)$ phase portrait can be labelled by the value $e_c$, which represent the eccentricity at the centre of the resonance. Since the eccentricity along individual orbits exhibits small variations, using $e_c$ as a label for phase portraits allows for a direct comparison of the theoretical phase portraits with the distributions of Hildas at different eccentricities of $e$. In Fig. \ref{aphi1}, we display phase portraits in the $(a,\phi_1)$ plane for $e_c=$0.1, 0.2, or 0.3, corresponding to the three $e$ values that we considered. In each panel, the black curves denote the resonant islands and the red curve denotes the resonant boundary between the libration and circulation regions; and the dashed and solid curves refer to the cases before and after the jump of Jupiter, respectively, i.e. the original and new resonances as we defined above. And, the dots represent the initial test Hildas.

For better visualization of the `shift of resonant centre' scenario, we present the actual phase portraits of the 3:2 Jovian resonance in the $(a,\phi_1)$ plane. Note that this plane is equivalent to the `canonical' $(\Phi_1,\phi_1)$ plane but is more intuitive. The $(a,\phi_1)$ phase portrait can be labeled by the value of $e_c$, which represents the eccentricity at the centre of the resonance. Since the eccentricity along individual orbits exhibits small variations, using $e_c$ as a label for phase portraits allows for a direct comparison of the theoretical phase portraits with the distributions of Hildas at different eccentricities of $e$. In Fig. \ref{aphi1}, we display phase portraits in the $(a,\phi_1)$ plane for $e_c=0.1$, 0.2, and 0.3, corresponding to the three $e$ values we considered. In each panel, the black curves denote the resonant islands, and the red curve marks the resonant boundary between the libration and circulation regions. The dashed and solid curves refer to the cases before and after the jump of Jupiter, respectively -- that is, the original and new resonances. The dots stand for the test Hildas.

Now it becomes much easier to illustrate how the resonant centre shifts and the associated consequences. As shown in Fig. \ref{aphi1}, after the shift of the Jovian 3:2 resonance, only the test Hildas coloured in blue can survive. These objects are located in the overlapping region between the original resonance (inside the dashed red boundary) and the new resonance (inside the solid red boundary). While the yellow Hildas, which are outside the field of the new resonance, would escape. We can then show the differences among the test Hildas with various eccentricities by examining the central resonant island enclosed by the innermost dashed/solid black curve, where the original/new resonant centre is located. In the case of $e_c=0.1$ (Fig. \ref{aphi1}, top panel), all the blue Hildas are far from the region near the new resonant centre, i.e. outside the innermost solid black curve. One can take the blue Hilda plotted on the outermost solid black curve as an example. Since a Hilda must evolve along this curve to conserve the Hamiltonian (or equivalently, the energy), it will experience large variations in the angle $\phi_1$. Keep in mind that, $\phi_1$ is actually the resonant angle $\sigma$. Therefore, the surviving Hildas with $e\sim0.1$ would end with large resonant amplitudes. However, in the cases of $e_c=0.2$ and 0.3 (Fig. \ref{aphi1}, middle and bottom panels), one can observe that there are blue Hildas located on or inside the individual innermost solid black curves, i.e. very close to the new resonant centre. These more eccentric objects could experience smaller variations in the angle $\phi_1$ ($=\sigma$) and thus have much lower resonant amplitudes at the end of the evolution.

%________________________________________________________________________________________________________________________________________________________

\section{Orbit and mass of the FFP}
\label{orbitandmass}

As mentioned at the end of Sect. 2.1, an incoming FFP into the Solar System was clearly on a hyperbolic orbit rather than a parabolic one. Nevertheless, the key mechanism is a jumping Jupiter scenario, which can be triggered by the FFP with various types of unbounded orbits. Specifically, we aim to achieve Jupiter's outward migration at a distance of $\Delta a_J\sim0.12$ AU on a timescale of $\sim10$ yr. For the sake of simplicity, we used a parabolic orbit before, but a hyperbolic orbit would produce the same effect. In this section, we will perform additional calculations to determine the FFP's potential hyperbolic orbit, as well as its mass.

\begin{figure}
 \hspace{0cm}
  \centering
  \includegraphics[width=8cm]{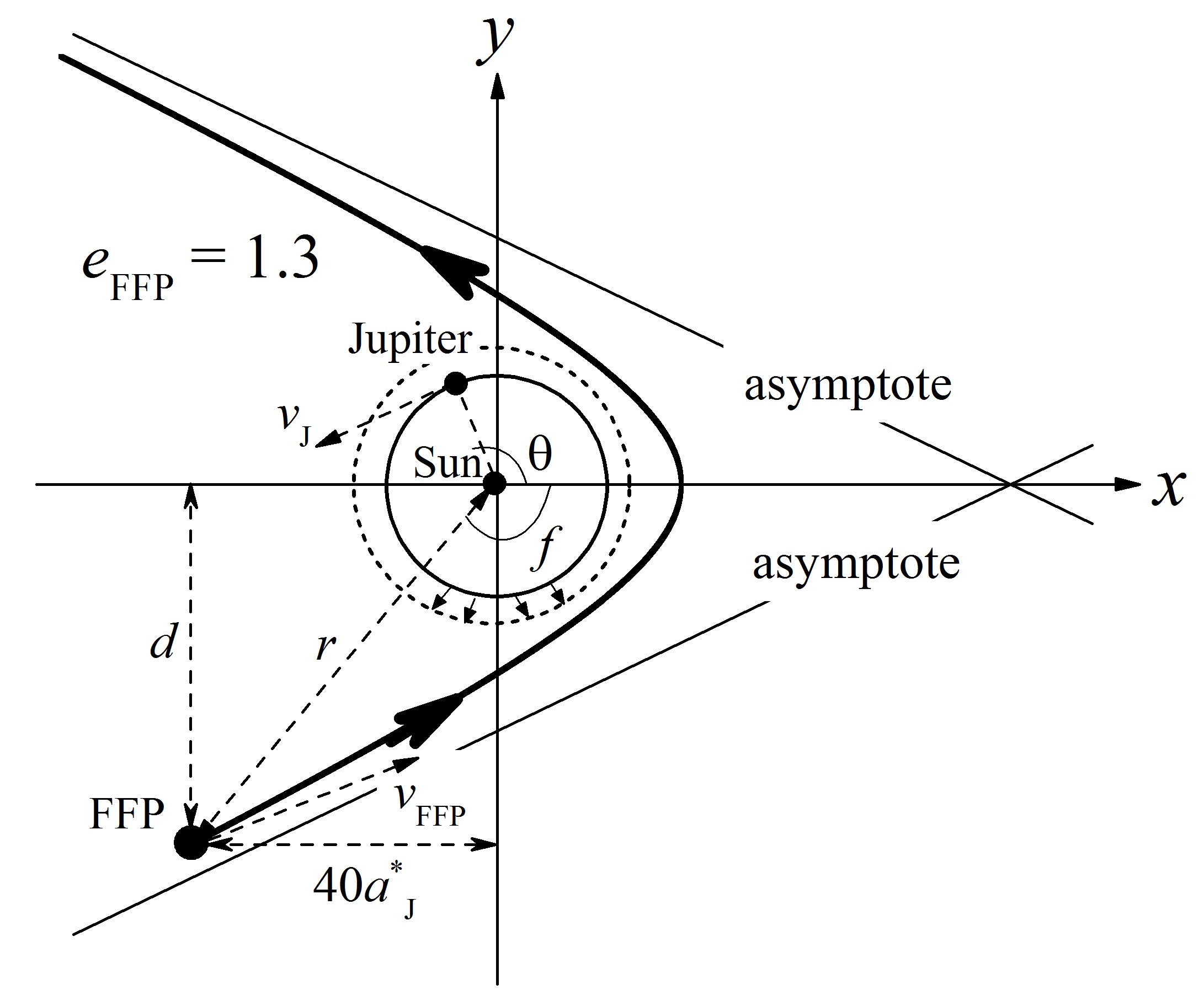}
  \caption{Schematic diagram of the hyperbolic flyby of an FFP and the resulting outward jump of Jupiter. Jupiter starts on the circular orbit (solid curve) around the Sun, at a distance of $a_J^{\ast}$ and with a phase angle of $\theta$. The FFP approaches on a hyperbolic coplanar orbit from a location at $(-40a_J^{\ast}, -d)$ ($d>0$), with an eccentricity of $e_{FFP}>1$. After a strong interaction between these two planets near the FFP's perihelion on the positive $x$-axis, Jupiter moves from its original orbit (solid curve) to an expanded one (dashed curve), and the hyperbolic FFP leaves the Solar System permanently.}
  \label{ffp}
\end{figure}

Similar to \citet{Li2023}, we begin by analysing the dynamics of a coplanar three-body system comprising the Sun of mass $M_{\odot}$, Jupiter of mass $m_J$, and an FFP of mass $m_{FFP}$, except for the FFP being assumed to be on a hyperbolic orbit characterised by an eccentricity of $e_{FFP}>1$. The initial configuration and evolution of this three-body system are shown in Fig. \ref{ffp}. In the heliocentric coordinate system $(x, y)$, Jupiter is initially placed on a circular orbit with a radius of $a_J^{\ast}$. Correspondingly, its initial velocity is calculated as:  
\begin{equation}
v_J(0)=\sqrt{\frac{G(M_{\odot}+m_J)}{a_J^{\ast}}}.
\label{vJ}
\end{equation}
Here and henceforth, the bracketed digit 0 indicates the initial parameter values at time $t=0$. Then the initial positions and velocities of Jupiter are given by:
\begin{eqnarray}
x_J(0)&=&a_J^{\ast}\cos\theta, ~~~~~~~~~~ y_J(0)=a_J^{\ast}\sin\theta,\nonumber\\
\dot{x}_J(0)&=&-v_J(0)\sin\theta,~~~~~\dot{y}_J(0)=v_J(0)\cos\theta,
\label{initialJ}
\end{eqnarray}
where the initial phase angle $\theta$ is measured counterclockwise from the  positive $x$-axis. Note that the scale of this three-body system is defined by Jupiter's initial heliocentric distance $a_J^{\ast}$. Therefore, all distances will be expressed in terms of $a_J^{\ast}$. While we have taken $a_J^{\ast}$ to be 5.2 AU as we did in \citet{Li2023}, it's worth mentioning that this value could vary if other orbital configurations are to be considered. 

The FFP is coming from a specific location of $(-40a_J^{\ast}, -d)$ ($d>0$) and it follows a hyperbolic orbit with an eccentricity of $e_{FFP}$ ($>1$). As we will illustrate below, given a value of $e_{FFP}$, the impact parameter $d$ controls the interaction between Jupiter and the FFP when they approach each other close to the FFP's perihelion on the positive $x$-axis. The hyperbolic orbit of the FFP is described by 
\begin{equation}
r=\frac{p}{1+e_{FFP}\cos f},
\label{radial}
\end{equation}
where $r$ represents the radial distance, $f$ is the true anomaly, and $p$ is the semilatus rectum. Then the initial positions and velocities of the FFP can be written as
\begin{eqnarray}
x(0)\!\!\!\!&=&\!\!\!\!-40a_J^{\ast}, ~~~~~~~~~~~~~~~~~~~~ y(0)=-d,\nonumber\\
\dot{x}(0)\!\!\!\!&=&\!\!\!\!-\sqrt{{\tilde{\mu}}/{p}}\sin f(0),~\dot{y}(0)=\sqrt{{\tilde{\mu}}/{p}}~(e_{FFP}\!+\!\cos f(0)),
\label{initialFFP}
\end{eqnarray}
where $\tilde{\mu}=G(M_{\odot}+m_{FFP})$.

Consider the energy integral of the hyperbolic motion
\begin{equation}
K=\frac{1}{2}v^2-\frac{\tilde{\mu}}{r}=\frac{\tilde{\mu}}{2a},
\label{energy}
\end{equation}
and the angular momentum integral
\begin{equation}
r^2\dot{f}=\sqrt{\tilde{\mu} p},
\label{angularmomentum}
\end{equation}
where the FFP's velocity $v$ satisfies $v^2=\dot{r}^2+r^2\dot{f}^2$ and the semi-major axis $a=p/(e_{FFP}^2-1)$. We can then have
\begin{equation}
\dot{r}^2=\frac{\tilde{\mu}}{a\cdot r^2}[(r+a)^2-a^2e_{FFP}^2].
\label{dot_r2}
\end{equation}
If we define a pseudo mean-motion frequency $\nu$ for the FFP as
\begin{equation}
{\nu}^2 a^3=\tilde{\mu}=G(M_{\odot}+m_{FFP}),
\end{equation}
by integrating Eq. (\ref{dot_r2}) we can write
\begin{equation}
{\nu}\mbox{d}t=\frac{r\mbox{d}r}{a\sqrt{(r+a)^2-a^2e_{FFP}^2}}.
\label{nudt}
\end{equation}
Now, introducing an auxiliary variable $F$ as
\begin{equation}
r=a(e_{FFP} \cosh{F}-1),
\label{rcosh}
\end{equation}
where $\cosh$ is the `hyperbolic cosine' function. By substituting for $r$ from Eq. (\ref{rcosh}) into Eq. (\ref{nudt}), the integration yields the Keplerian equation for the hyperbolic orbit:
\begin{equation}
e_{FFP}\sinh{F}-F={\nu}(t-t^{\ast}),
\label{rsinh}
\end{equation}
where the integral constant $t^{\ast}$ represents the time of the FFP's perihelion passage located on the positive $x$-axis.

Next, we consider the close encounter between Jupiter and the FFP. First, assuming such a strong interaction happens around the FFP's perihelion, we require that Jupiter also reaches the positive $x$-axis at the very moment of $t=t^{\ast}$. Accordingly, we can determine Jupiter's initial phase angle to be
\begin{equation}
\theta^{\ast}=\frac{t^{\ast}\pmod{T_J}}{T_J}\cdot(-2\pi),
\label{sita}
\end{equation}
where $T_J\approx11.86$ yr is the orbital period of Jupiter for $a_J=a_J^{\ast}=5.2$ AU. Secondly, when Jupiter and the FFP are both positioned on the positive $x$-axis, their coordinates are given as $(a_J^{\ast}, 0)$ and $(a(e_{FFP}-1), 0)$, respectively. To ensure a close encounter but prevent a collision between these two planets, the following condition must be met:
\begin{equation}
a(e_{FFP}-1)>a_J^{\ast},
\end{equation}
This condition can be equivalently expressed as:
\begin{equation}
p>(e_{FFP}+1)a_J^{\ast}.
\label{condition}
\end{equation}
We can determine the value of $p$ using the initial positions of the FFP through 
\begin{eqnarray}
p&=&r(0)\cdot(1+e_{FFP}\cos f(0))=r(0)+e_{FFP}\cdot x(0) \nonumber\\
&=&\sqrt{(x(0))^2+(y(0))^2}+e_{FFP}\cdot x(0),
\label{pvalue}
\end{eqnarray}
By substituting the values of $x(0)$ and $y(0)$ from Eq. (\ref{initialFFP}) into Eqs. (\ref{pvalue}), Ineq. (\ref{condition}) becomes 
\begin{equation}
\sqrt{(40a_J^{\ast})^2+d^2}-e_{FFP}\cdot40a_J^{\ast}>(e_{FFP}+1)a_J^{\ast}.
\label{dvalue}
\end{equation}
This way, when the impact parameter $d$ exceeds a certain critical value, which of course depends on $e_{FFP}$, the FFP will always remain outside Jupiter's orbit and drive Jupiter to migrate via a close encounter, as visualised in Fig. \ref{ffp}.

\begin{figure}
 \hspace{0cm}
  \centering
  \includegraphics[width=9cm]{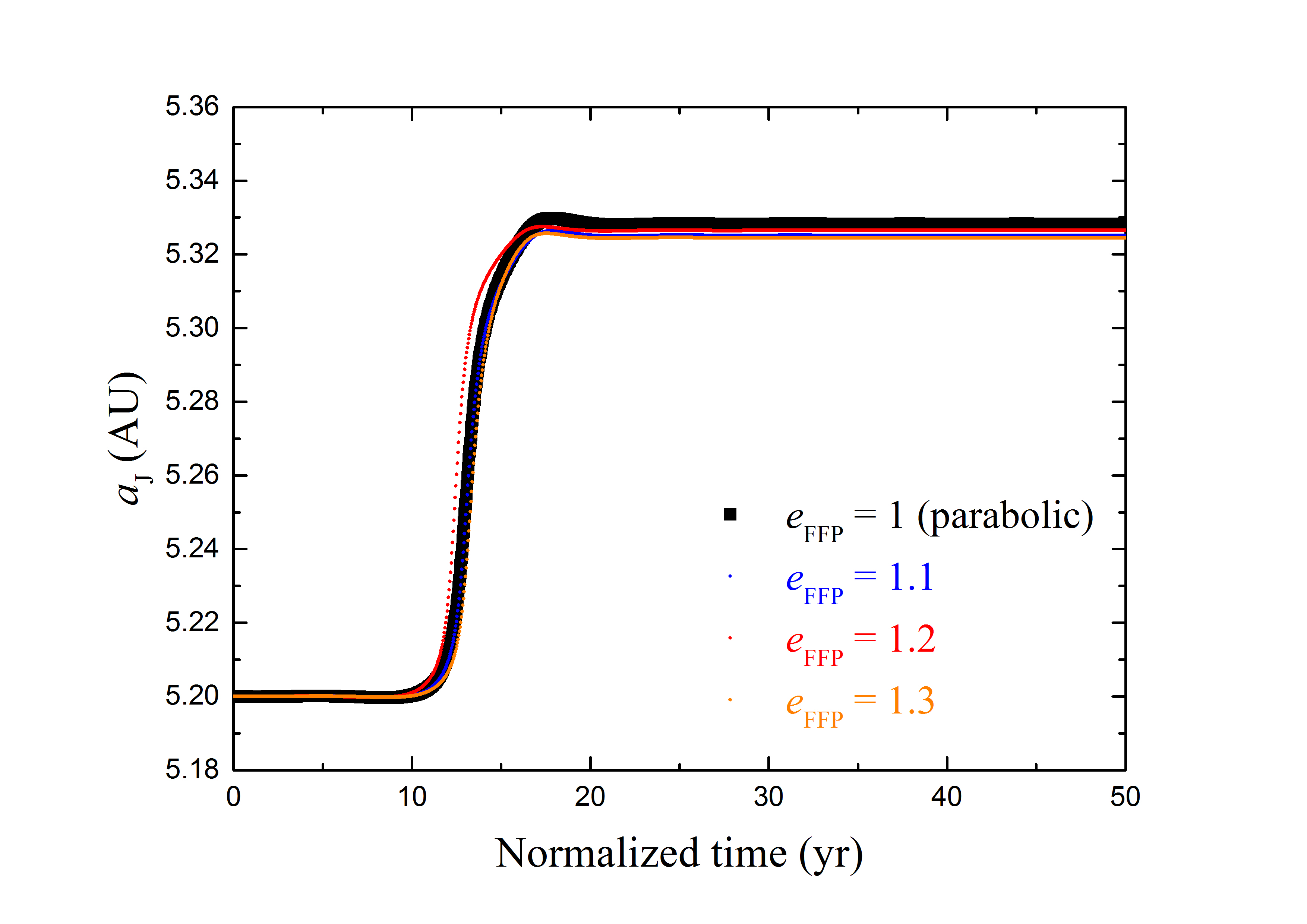}
  \caption{The temporal evolution of Jupiter's semi-major axis $a_J$ as a consequence of the FFP flyby. The FFP follows a parabolic orbit with an eccentricity of $e_{FFP}=1$ (black), or a hyperbolic orbit with $e_{FFP}=1.1$ (blue), 1.2 (red), and 1.3 (orange). The FFP has a mass of Jupiter mass (i.e. $m_{FFP}=m_J$) and an inclination of zero (i.e. $i_{FFP}=0$).}
  \label{ffpEcc}
\end{figure}

\begin{figure}
  \centering
  \begin{minipage}[c]{0.5\textwidth}
  \centering
  \hspace{0cm}
  \includegraphics[width=9cm]{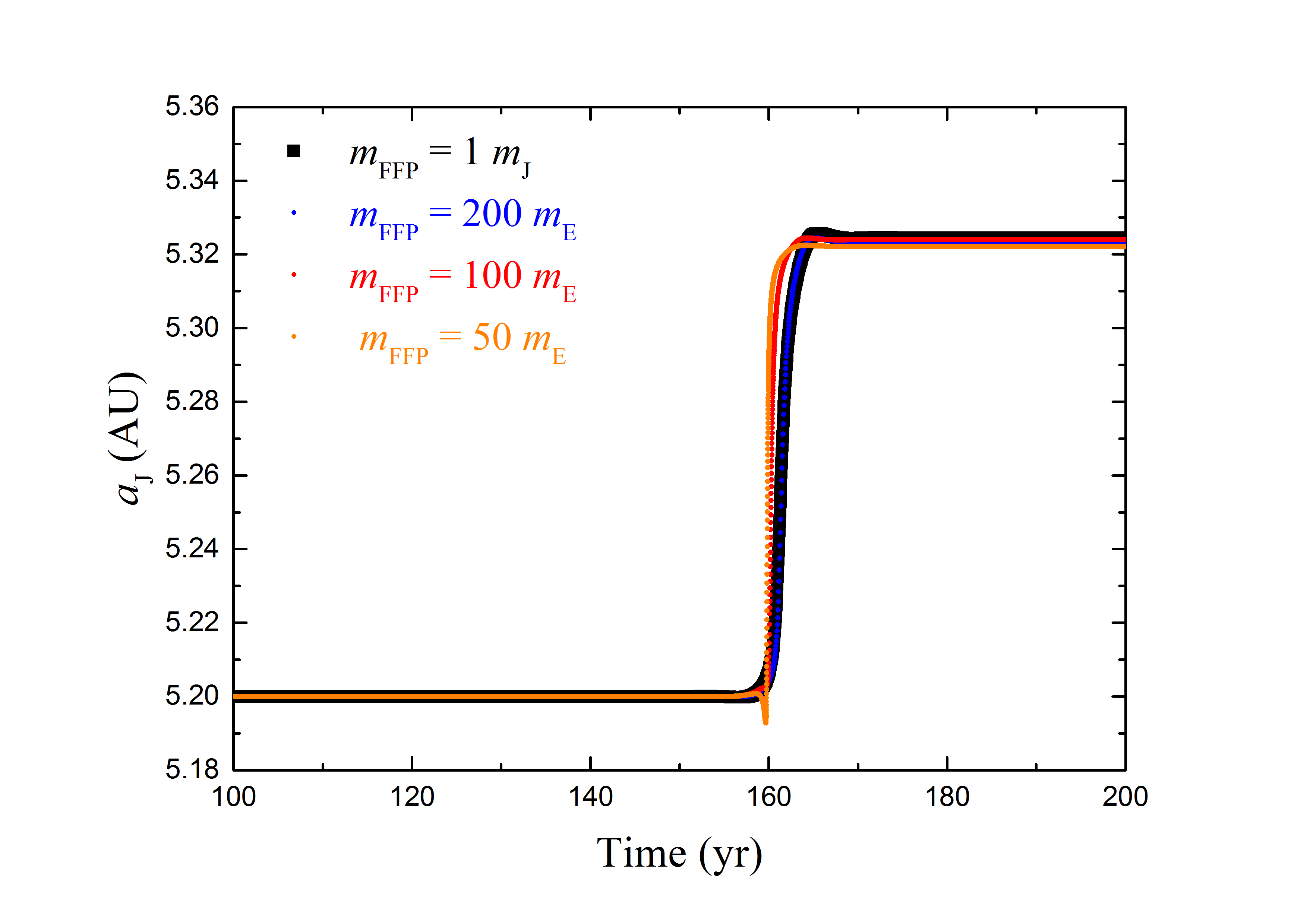}
  \end{minipage}
  \begin{minipage}[c]{0.5\textwidth}
  \centering
  \hspace{0cm}
  \includegraphics[width=9cm]{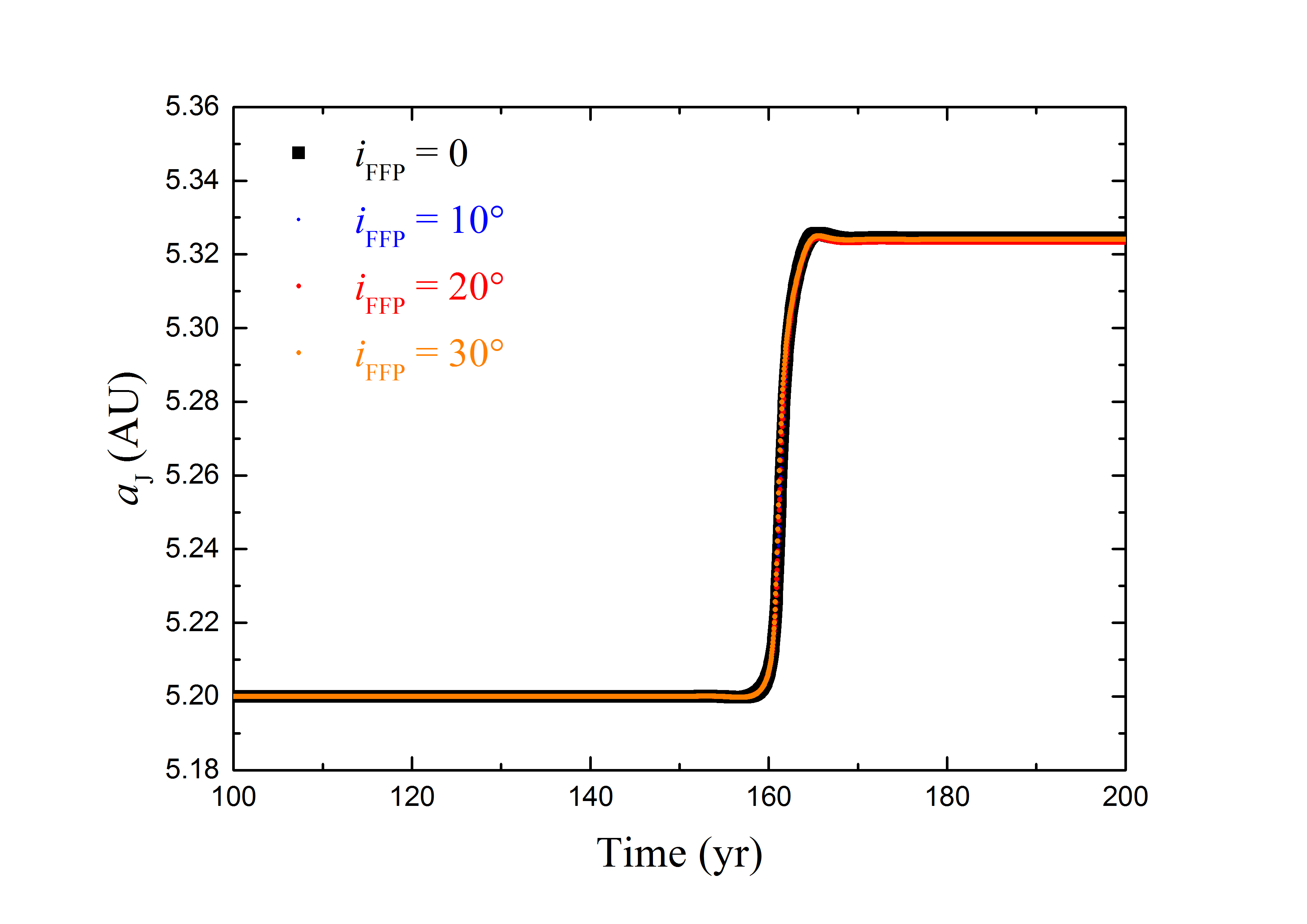}
  \end{minipage}
 \caption{Similar to Fig. \ref{ffpEcc} but only for the hyperbolic case of $e_{FFP}=1.3$. (Top panel) The FFP has a mass of $m_{FFP}=m_J$ (black), 200 $m_E$ (blue), 100 $m_E$ (red), and 50 $m_E$ (orange), where $m_J$ and $m_E$ denote Jupiter's mass and Earth's mass, respectively. (Bottom panel) The FFP has an inclination of $i_{FFP}=0$ (black), $10^{\circ}$ (blue), $20^{\circ}$ (red), and $30^{\circ}$ (orange). Note that the black curves in both panels correspond to the reference case of $m_{FFP}=m_J$ and $i_{FFP}=0$, which is identical to the case of $e_{FFP}=1.3$ depicted in Fig. \ref{ffpEcc} (orange curve).}
 \label{ffpIncMass}
\end{figure}

To account for the number asymmetry of Jupiter Trojans (in \citet{Li2023}) and the resonant amplitude distribution of Hildas (in this paper), the key mechanism is Jupiter's outward jump, as indicated by the black curve in Fig. \ref{ffpEcc}. This temporal evolution of Jupiter's semi-major axis $a_J$ was achieved by simulating the parabolic approach of an FFP with an eccentricity of $e_{FFP}=1$. Thus in the case of a hyperbolic FFP (i.e. $e_{FFP}>1$), our objective is to ensure that Jupiter can follow this specific $a_J$ evolution.

Considering that the interstellar object 1I/`Oumuamua has an eccentricity of 1.2, for the hyperbolic FFP, we choose three representative eccentricities: $e_{FFP}=1.1$, 1.2, and 1.3. For each value of $e_{FFP}$, we assign an appropriate impact parameter $d$, and the resulting jump of Jupiter is indicated by individual colourful curves in Fig. \ref{ffpEcc}. A comparison of these jumps with the one induced by a parabolic FFP (i.e. with $e_{FFP}=1$), as depicted by the black curve in this figure, reveals a remarkable similarity. Both the distances ($\sim 0.12$ AU) and timescales ($\sim 10$ yr) of $a_J$ variations are nearly identical between the parabolic and hyperbolic cases. This can nicely support the validity of our simplification in using a parabolic FFP. We note that the time $t^{\ast}$ of the FFP's perihelion passage would vary slightly with different values of $e_{FFP}$. For consistency, in Fig. \ref{ffpEcc}, time values have been normalized so that Jupiter starts to jump at the same time point.

For reference, we here provide the relative velocity during the encounter, assumed to occur around the timing of the FFP's perihelion passage. It is easy to determine the velocity of the FFP at perihelion. Setting $f=0$ in Eq. (\ref{radial}), we obtain the perihelion's radial distance of $r_p=p(1+e_{FFP})$. Substituting $r=r_p$, $a=p/(e^2_{FFP}-1)$ and the value of $p$ from Eq. (\ref{pvalue}) into Eq. (\ref{energy}), we get
\begin{eqnarray}
\dot{x}_{peri}\!\!\!\!&=&\!\!\!\!0~,\nonumber\\
\dot{y}_{peri}\!\!\!\!&=&\!\!\!\!(1+e_{FFP})\sqrt{{\tilde{\mu}}/{p}}\nonumber\\
\!\!\!\!&=&\!\!\!\!(1+e_{FFP})\sqrt{\frac{{G(M_{\odot}+m_{FFP})}}{{\sqrt{(40a_J^{\ast})^2+d^2}-e_{FFP}\cdot 40a_J^{\ast}}}}~,
\label{FFPperiV}
\end{eqnarray}
where the subscript `\textit{peri}' is the abbreviation for `perihelion'. Taking the case of $e_{FFP}=1.3$ as an example, to satisfy Ineq. (\ref{dvalue}), we chose $d=36.8 a_J^{\ast}$. Then, the FFP's perihelion traversing would have a velocity of (0, +19.59 km/s). Using Eqs. (\ref{vJ}) and (\ref{initialJ}), when Jupiter is also passing through the $x$-axis, it has a velocity of (0, +13.07 km/s). Accordingly, the relative velocity between these two planets during the encounter is about 6.52 km/s.

By now, we have adopted the FFP with a mass of $m_{FFP}=m_J$ and an inclination of $i_{FFP}=0$. We would like to further explore the possible ranges of $m_{FFP}$ and $i_{FFP}$ that can produce a similar Jupiter jump as displayed in Fig. \ref{ffpEcc}. Using the hyperbolic FFP with $e_{FFP}=1.3$ as an illustration, a series of additional tests have been carried out. The results are presented in Fig. \ref{ffpIncMass}. In both panels, the black curves depict the reference case of $m_{FFP}=m_J$ and $i_{FFP}=0$, i.e. the case of $e_{FFP}=1.3$ indicated by the orange curve in Fig. \ref{ffpEcc}. We find that, for the FFP with a mass as small as 50 $m_E$ and an inclination as high as $30^{\circ}$, the resulting Jupiter jump can also closely resemble that observed in the reference case. We would like to note that the top panel in Fig. \ref{ffpIncMass} suggests that an encounter with an FFP as massive as Jupiter can be as effective as one with a much smaller mass. Actually, this effectiveness is also influenced by the impact parameter $d$ associated to the FFP and the initial phase angle $\theta$ of Jupiter. As the FFP's mass $m_{FFP}$ decreases, a smaller value is chosen for $d$, ranging from 36.8 $a_J^{\ast}$ to $\sim36.7$ $a_J^{\ast}$, bringing the FFP's perihelion closer to Jupiter's orbit. Additionally, the angle $\theta$, which is around $\theta^{\ast}$, is adjusted accordingly to further reduce the relative distance between these two planets. As a result, the acceleration for Jupiter induced by a less massive FFP can be compensated for by varying $d$ and $\theta$.

Taken in total, we propose that the FFP that flew by the Solar System could possess a mass of $m_{FFP}$ at least approximately 50 Earth masses, an eccentricity of $e_{FFP}=1$-$1.3$, along with an inclination $i_{FFP}$ as large as $30^{\circ}$. Remarkably, all these diverse parameters can lead to nearly the same outward jump of Jupiter, that is a displacement of $\sim 0.12$ AU over a time span of $\sim 10$ yr. This specific Jupiter jump serves as a key mechanism for replicating the individual observational features in both the Jupiter Trojan and Hilda populations simultaneously.

%______________________________________________________________

\section{Conclusions and discussion}

In this paper, we demonstrate that a single FFP flyby, previously proposed to explain the number asymmetry of Jupiter Trojans \citep{Li2023}, can simultaneously replicate the resonant amplitude distribution observed in the Hilda asteroids. The FFP flyby could cause a rapid outward migration of the 3:2 Jovian resonance, where the Hildas are located. Consequently, many original Hildas with small resonant amplitudes, which should be the stable ones, would escape from the 3:2 resonance. The resonant orbits of survived Hildas exhibit two distinct patterns that match observations perfectly: (1) a lack of Hildas with resonant amplitudes $A<40^{\circ}$ at eccentricities $e<0.1$; (2) an almost complete absence of Hildas with $A<20^{\circ}$, regardless of $e$. In addition, we argue that even if the original Hildas arising from various formation models may have different initial $A$-distributions -- whether less excited, overexcited, or excited to the right degree -- in our FFP flyby scenario, these two resonant patterns of Hildas can be consistently present.

These results suggest that the same FFP flyby scenario used to explain the number asymmetry of Jupiter Trojans \citep{Li2023} could potentially reproduce the absence of the most stable Hildas with small resonant amplitudes, not just qualitatively but also quantitatively. Or at least, this scenario would not result in a resonant amplitude distribution of the Hildas that conflicts with current observations. Thus, our previously published FFP hypothesis regarding the Jupiter Trojans appears to be supported.  

%It is important to note that the fundamental mechanism involves a fast outward jump of Jupiter by $\sim0.12$ AU, while the FFP flyby near Jupiter's orbit serves as a potential trigger. Our calculation indicates that this FFP should have a minimum mass of about 50 Earth masses, an eccentricity varying from 1 to 1.3, or even larger, and an inclination reaching up to $30^{\circ}$ or higher. We remark that some authors have suggested a penetrating encounter at a certain inclination, such as much as $30^{\circ}$, similar to what was obtained herein. This is bound to alter the distribution of inclinations of the perturbed small body population \citep[e.g.][]{Moor23}. Upon examining the case of an FFP with a $30^{\circ}$ inclination, we find that the excitation of inclinations of the simulated Hildas would be less than $1^{\circ}$. This suggests that the inclination distribution of the Hildas would not be significantly affected in our FFP flyby scenario.

We note that in our hypothesis, the crucial step is a fast outward jump in Jupiter's semi-major axis by $\Delta a_J\sim0.12$ AU over a timescale of $\Delta t\sim10$ yr, while the FFP flyby near Jupiter's orbit serves as a potential trigger. Our calculation indicates that this FFP should have a minimum mass of about 50 Earth masses, an eccentricity varying from $e_{FFP}=1-1.3$, or even larger, and an inclination reaching up to $i_{FFP}=30^{\circ}$ or higher. We remark that some authors have suggested a penetrating encounter at a certain inclination, such as much as $30^{\circ}$, similar to what was obtained herein. This is bound to alter the distribution of inclinations of the perturbed small body population \citep[e.g.][]{Moor23}. Upon examining the case of an FFP with a $30^{\circ}$ inclination, we find that the excitation of inclinations of the simulated Hildas would be less than $1^{\circ}$. This suggests that the inclination distribution of the Hildas would not be significantly affected in our FFP flyby scenario.

\begin{figure}
 \hspace{0cm}
  \centering
  \includegraphics[width=9cm]{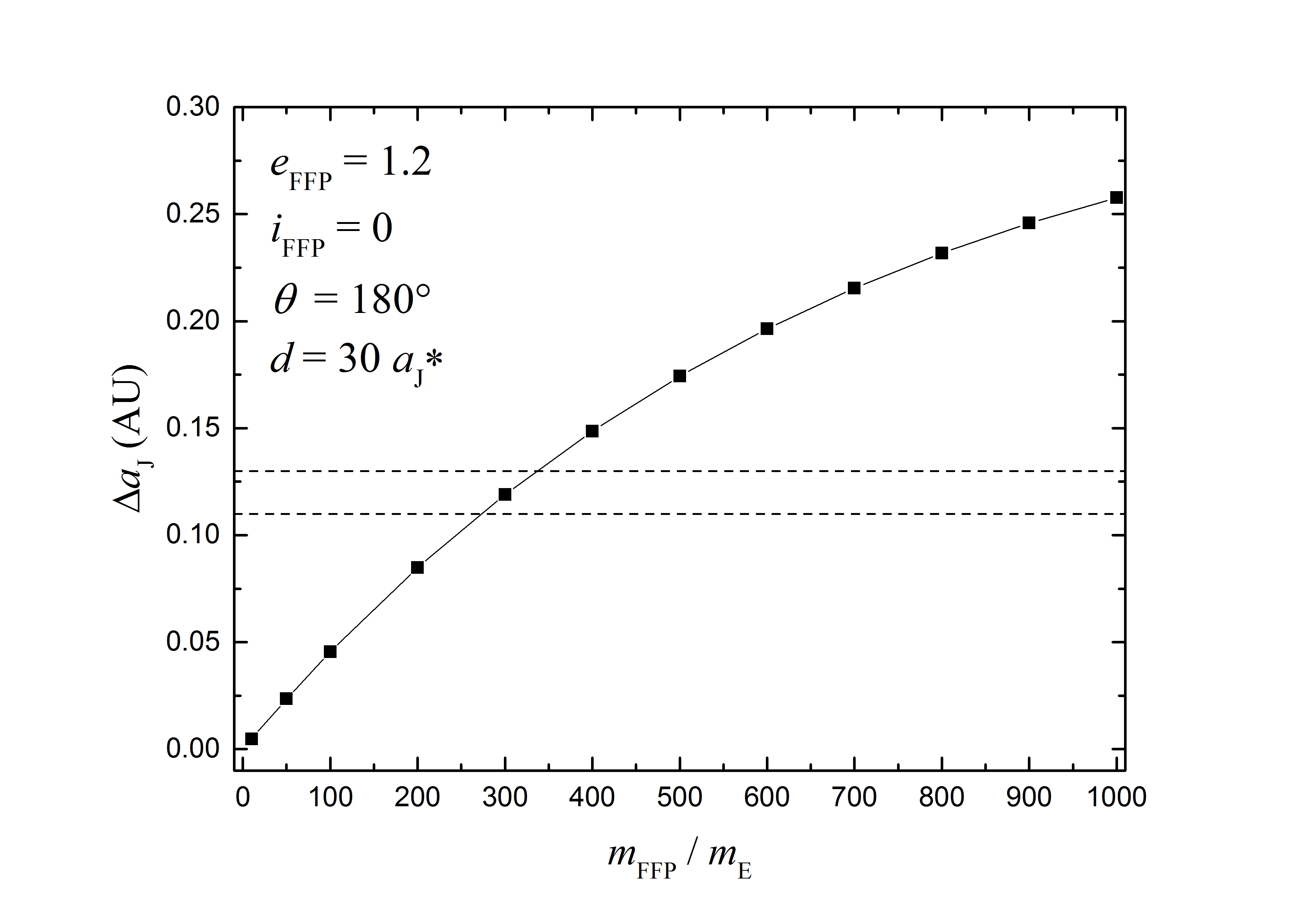}
  \caption{The variation of Jupiter's semi-major axis, $\Delta a_J$, against the mass of the FFP, $m_{FFP}$. The other free parameters of the FFP flyby scenario, $e_{FFP}$, $i_{FFP}$, $d$ and $\theta$, are kept fixed. For reference, two horizontal dashed lines are plotted at $\Delta a_J=0.11$ AU and 0.12 AU, respectively.}
  \label{mFFPDa}
\end{figure}

One might wonder whether the crucial step, namely the fast jump of Jupiter, can alone account for the resonant amplitude distribution of the Hildas. To address this, we repeated the experiment from Sect. 2 with $e_{FFP}=1$, keeping the dynamical model the same but removing the FFP's direct influence on the Hildas. This modification does not affect the evolution of Jupiter, resulting in exactly the same jump of Jupiter with $\Delta a_J\sim0.12$ AU and $\Delta t\sim10$ yr. The results show that similar resonant amplitude distributions can be achieved for the test Hildas with $e=0.2$ and $e=0.3$. However, none of the test Hildas with $e=0.1$ survive. The reason could be that the FFP’s direct gravitational influence on the Hildas should be included to maintain the consistent angular motion of both Jupiter and the Hildas. Unlike the classical planet migration model, where particles are captured by the planet's resonance \citep{Malh1995, Li2014a, Li2014b}, the test Hildas were initially inside the 3:2 Jovian resonance. Due to the jump of Jupiter, they are displaced from the original resonance center in the $a$-direction. In the azimuthal direction, if the FFP only affects Jupiter, rather than the Jupiter+Hilda system as a whole, it would induce an angular acceleration of Jupiter by about $10^{\circ}$ right after the FFP leaves, while having no effect on the Hildas. This leads to additional variation in the resonant configuration, affecting the resonant angle beyond just $a$. As shown in Fig. \ref{aphi1} (where $\phi_1$ represents the resonant angle), for the test Hildas with larger eccentricities of $e=0.2$ and 0.3, the overlap between the original and new resonance regions is substantial. Therefore, the variation of the resonant angle may have minimal impact on the retention of these test Hildas. However, for the test Hildas with $e=0.1$, the overlapping region is much smaller. Then those objects initially in the resonance overlap zone could all be expelled from the resonance due to the resonant angle variation. To further validate our conclusion that a similar resonant amplitude distribution of the Hildas can be produced when the FFP causes the same jump of Jupiter, we performed an experiment with an FFP flyby at $e_{FFP}=1.1$. As shown in Fig. \ref{ffpEcc}, this also results in the same jump of Jupiter as observed at $e_{FFP}=1$. We find that the final distributions of test Hildas for $e=0.1$, 0.2, and 0.3 are consistent with those presented in Fig. \ref{randomEcc} (b). Although further studies are needed to explore the direct effect of the FFP on the Hildas in detail, the results obtained in this paper are robust. We propose that to reproduce the resonant amplitude distribution of the Hildas, the jump of Jupiter might essentially be related to an FFP passing. Since the direct gravitational influence of the FFP on the Hildas may not be neglected, the jump of Jupiter and the FFP flyby should not be treated as separate events. This may further emphasize the importance of the FFP flyby scenario, and it is plausible that an FFP did intrude into our Solar System, as conjectured by \citet{Doul19}. Nevertheless, another perturber capable of producing a similar gravitational effect in the Jupiter region could also be a possibility.

The FFP flyby scenario examined in this paper focuses on a specific jump of Jupiter, with a semi-major axis change of $\Delta a_J\sim0.12$ AU over a timescale of $\Delta t\sim10$ yr. This indicates a very fast migration with a speed of ${\Delta a_J}/{\Delta t}=0.012$ AU/yr. It might be questioned whether a significantly slower speed would affect this proposed scenario. The answer is that the speed of Jupiter's jump is very relevant. To illustrate this, we performed an additional numerical simulation using a slower jump, with $\Delta a_J=0.5$ AU and $\Delta t\sim 1/3\times 10^4$ yr, giving a jump speed of $1.5\times10^{-4}$ AU/yr. This slower speed is derived from the planet instability model proposed by \citet{Nesv13} and was later used by \citet{Li2023a} to explain the number asymmetry of Jupiter Trojans. With such a slower jump, the quasi-integral $\Phi_2$ (see Eq. (\ref{Eq1})) would become an approximate adiabatic invariant of the non-autonomous flow, remaining conserved for test Hildas. Consistent with this theoretical argument, our numerical simulation reveals that test Hildas would just `follow' the change of the position of the 3:2 Jovian resonance, gradually migrating outward. At the end of the simulation, many of them remain in the librational domain. Although their eccentricities would decrease due to the conservation of the Jacobi constant \citep{Gomes97, Li2014a}, their resonant amplitudes would be nearly unchanged. Therefore, this slower jump of Jupiter cannot reproduce the observed distribution of Hildas in the resonant amplitude. We note that if $\Delta a_J$ is reduced to the previously considered value of 0.12 AU but $\Delta t$ is kept at $\sim 1/3\times 10^4$ yr, the results would be similar, as the jump becomes even slower at a speed of about $3.6\times10^{-5}$ AU/yr. At this point, the FFP flyby and Jupiter's fast jump scenario seem to be the preferred mechanism, as it can simultaneously account for both the number asymmetry of Jupiter Trojans and the resonance amplitude distribution of Hildas. To reproduce the observed distribution of Hildas, we suggest that the speed of Jupiter's jump, ${\Delta a_J}/{\Delta t}$, has to exceed $1.5\times10^{-4}$ AU/yr. However, refining the lowermost possible value of ${\Delta a_J}/{\Delta t}$ only for the Hildas is of limited significance, we must also take into account the relevant dynamical processes contributing to the number asymmetry of Jupiter Trojans.

Let us recall that the fast outward jump of Jupiter in the FFP flyby scenario was initially proposed to replicate the L4-to-L5 number ratio of Jupiter Trojans, precisely matching the observed ratio of $\sim 1.6$ \citep{Li2023}. To achieve this exact number ratio, we determined $\Delta a_J$ to be $\sim0.12$ AU, while $\Delta t$, the timescale of the jump, is always at the level of 10 yr. This timescale is constrained by the duration of the perihelion passage of the FFP, which can hardly change according to the possible orbit and mass of the FFP. Although the primary focus of this paper is to investigate the effect of this specific jump of Jupiter on the distribution of Hildas, we will also make some discussion on how our choice of the parameters in the FFP flyby scenario can be adjusted.

In fact, for the Hildas, $\Delta a_J$ could take a value within a range around 0.12 AU. To reproduce the observed resonant distribution at $e=0.1$, according to the phase portraits shown in Figs. \ref{width} and \ref{aphi1}, the shift of the resonance centre in the positive $a$-direction should meet two conditions: (1) exceeding the half-width of the resonance, so that there are no initial Hildas near the new resonant centre, and (2) not exceeding the total width of the resonance, ensuring that the original and new resonance regions can overlap, thereby allowing for the retention of initial Hildas. Given that the resonance width at $e=0.1$ is about 0.18 AU (see Fig. \ref{width}), this yields a range of 0.09 AU $<\Delta a_J<$ 0.18 AU. However, this represents only a theoretical limit of $\Delta a_J$ for Hildas. It is important to remember that the FFP flyby scenario is primarily employed to explain the L4-to-L5 number ratio of Jupiter Trojans. According to \citet{Li2023}, if we choose $\Delta a_J=0.12\pm0.01$ AU, the resulting number ratio would be about 1.4-1.9, which is within the observationally possible range \citep{jewi04, naka08, grav11, grav12, slyu13}. This $\Delta a_J$ range appears more reasonable and satisfies the two conditions required for Hildas with $e=0.1$. As shown in Fig. \ref{aphi1}, for Hildas with $e=0.1$, 0.2, and 0.3, the displacement of the new resonance zones (indicated by solid black lines), which is only about $\pm 0.01$ AU, is much smaller than the width of the overlapping region of the original and new resonances (i.e. between the dashed and solid red curves). Thus, when $\Delta a_J$ varies within the range of 0.11-0.13 AU, the obtained results from our studies with $\Delta a_J=0.12$ AU for the distribution of Hildas should be largely unaffected.

As illustrated in Sect. 4, to achieve a specific jump in Jupiter's semi-major axis, $\Delta a_J$, we can deduce the possible parameters of the FFP flyby event, including the mass $m_{FFP}$, eccentricity $e_{FFP}$, inclination $i_{FFP}$, impact parameter $d$ for the FFP, and the initial phase angle $\theta$ of Jupiter. To produce the same $\Delta a_J$, these five free parameters are indeed coupled. Changing one parameter will alter the perturbation of the FFP on Jupiter and consequently change the value of $\Delta a_J$. For instance, reducing the mass $m_{FFP}$, one has to decrease the impact parameter $d$ to maintain an equal gravitational acceleration on Jupiter. Therefore, it is worth estimating the probability that a specific jump of Jupiter, with the assumed $\Delta a_J$, can happen as a percentage of the volume in the corresponding parametric space. Although this requires much more work to do later, here we will simply illustrate how $\Delta a_J$ changes by altering one of the five free parameters, while keeping the other four fixed. In this example, we alter the FFP's mass, $m_{FFP}$, while fixing $e_{FFP}=1.2$, $i_{FFP}=0$, $\theta=180^{\circ}$, and $d=30a_J^{\ast}$ (where $a_J^{\ast}$ is Jupiter's initial semi-major axis before the jump). We then plot $\Delta a_J$ against $m_{FFP}$ for $m_{FFP}$ ranging from 10 $m_{E}$ to 1000 $m_{E}$ in Fig. \ref{mFFPDa}. Based on the earlier discussion, we consider $\Delta a_J$ to fall within a small range around 0.12 AU, i.e. 0.11-0.13 AU. From Fig. \ref{mFFPDa}, we can infer that the mass of the FFP ranges between 277 $m_{E}$ and 340 $m_{E}$ (equivalent to 0.9-1.1 $m_{J}$), which seems capable of explaining the unusual patterns in both the Jupiter Trojan and Hilda populations. It can be seen that, in the range of $m_{FFP} < 340 m_{E} \sim 1.1 m_{J}$ (i.e. below the upper horizontal dashed line), $\Delta a_J$ is nearly proportional to $m_{FFP}$. This can be understood theoretically: if Jupiter's semi-major axis increases, it implies that Jupiter's orbital energy also increases, and this gravitational acceleration is proportional to the mass of the perturbing FFP. We finally note that the variation of $\Delta a_J$ with respect to $m_{FFP}$ is directly related to the variation of the speed $\Delta a_J/\Delta t$ of Jupiter's jump. Since changing $m_{FFP}$ hardly affects the duration of the FFP's rapid perihelion passage on an unbound orbit, that is, the period of its influence on Jupiter, the jump of Jupiter consistently occurs over a timescale of $\Delta t\sim10$ yr, as noted in Sect. 4. Therefore, the speed of Jupiter's jump can be described by $\sim \Delta a_J/$(10 yr), which is proportional to $\Delta a_J$.

In the future, once the L4-to-L5 number ratio of Jupiter Trojans is refined, our investigation process can be readily repeated, and the $\Delta a_J$ value (equivalent to the speed of Jupiter's jump) would be adjusted to determine the parameters of the FFP flyby scenario. It is evident that, as long as the L4-to-L5 number ratio is a fixed value rather than a range, one can infer that a specific jump of Jupiter is required, meaning the FFP should exert a proper perturbation. This implies that the parameters of the FFP flyby scenario are intrinsically specific, yet they are coupled and can vary in combination. Similarly, in a recent publication, \citet{Pfalzner24} demonstrated that to reproduce all the peculiar features of trans-Neptunian objects, the mass and orbit of a star that flew by the outer Solar System are also specific. We believe this represents a standard approach to inferring the parameters of the evolution model of the Solar System based on observational evidence.

Within the Hilda population, there are two asteroid groups around the largest objects, (1911) Schubart and (153) Hilda, known as the Schubart and Hilda families, respectively. Originating as fragments disrupted from a parent body, family members share similar orbits after the collision event. During the later long-term evolution, their eccentricity dispersion could be evolving over time due to the Yarkovsky effect. For the Schubart family, an estimated age of approximately 1.7 Gyr is suggested to match the observed eccentricity dispersion among its members \citep{Broz08}. If the epoch of the FFP flyby could be well determined, we may further explore the impact of this event on the Schubart and Hilda families. Nevertheless, it's noteworthy that most members in these two families are small enough to be affected by the Yarkovsky effect, while the large Hilda asteroids considered in this paper (as depicted in Fig. \ref{all} (a)) are not. Therefore, we believe that the presence of genetic families among the Hilda population would hardly affect our FFP flyby scenario for the distinct resonant patterns of Hildas, but it still holds potential to refine our FFP model in future studies.

In addition to affecting Jupiter Trojans and Hildas, the FFP flyby scenario proposed in this research may also influence other asteroid populations. As discussed in \citet{Carr15}, there are dynamically stable regions in the Cybele asteroid region that are characterized by a rather low number density of asteroids. The Cybele asteroids are positioned adjacent to and exterior of the main belt, with semi-major axes of $a\approx3.3$-3.7 AU. As non-resonant objects, the influence of the FFP flyby on them should be primarily driven by the direct perturbation from the FFP, rather than by the shift of the Jovian resonance(s) associated with the jump of Jupiter. We simulated the evolution of test Cybeles by considering an FFP flyby. Consistent with the study in Sect. 2, the incoming FFP follows the same parabolic orbit but varies in mass. Initially, the test Cybeles are uniformly spaced between 3.3 and 3.7 AU, with eccentricities $e$ randomly selected between 0 and 0.3. At the end of the 1 Myr simulation, we found that if the FFP's mass is less than a Jupiter mass (i.e. $m_{FFP} < m_J$), over 80\% of the test Cybeles can survive, with little change in the distributions of their $a$ and $e$. However, when $m_{FFP}$ exceeds $5m_J$, the outermost Cybeles asteroids at $a\approx3.7$ AU are strongly depleted. A similar issue of low number density is also observed in the main belt asteroids at high inclinations \citep{Carr11}, and a scenario involving an FFP intruding into the Solar System on an inclined orbit by as high as $30^{\circ}$ could be the reason. Future work within the framework of the FFP flyby scenario will investigate the low number density issues in these asteroid populations.

\section*{Funding and competing interests}

This work was supported by the National Natural Science Foundation of China (Nos. 12473061, 11973027, 11933001, 12150009), and National Key R\&D Program of China (2019YFA0706601). And part of this work was also supported by a Grant-in-Aid for Scientific Research (20H04617). The authors would like to express their thanks to the reviewers, Prof. Valerio Carruba, Prof. Rudolf Dvorak, and another anonymous referee for the valuable comments that helped to considerably improve the manuscript.

\printcredits

%\appendix
%\section{My Appendix}
%Appendix sections are coded under \verb+\appendix+.

\section*{Declaration of competing interest}
The authors declare that they have no known competing financial interests or personal relationships that could have appeared to influence the work reported in this paper.

\section*{Data Availability}

The data underlying this article are available in the article.

\bibliographystyle{cas-model2-names}

\bibliography{cas-refs}


%Author biography without author photo.
%Author biography. Author biography. Author biography.
%Author biography. Author biography. Author biography.
%Author biography. Author biography. Author biography.
%Author biography. Author biography. Author biography.
%Author biography. Author biography. Author biography.
%Author biography. Author biography. Author biography.
%Author biography. Author biography. Author biography.
%Author biography. Author biography. Author biography.
%Author biography. Author biography. Author biography.

%\bio{figs/cas-pic1}
%Author biography with author photo.
%Author biography. Author biography. Author biography.
%Author biography. Author biography. Author biography.
%Author biography. Author biography. Author biography.
%Author biography. Author biography. Author biography.
%Author biography. Author biography. Author biography.
%Author biography. Author biography. Author biography.
%Author biography. Author biography. Author biography.
%Author biography. Author biography. Author biography.
%Author biography. Author biography. Author biography.

%\bio{figs/cas-pic1}
%Author biography with author photo.
%Author biography. Author biography. Author biography.
%Author biography. Author biography. Author biography.
%Author biography. Author biography. Author biography.
%Author biography. Author biography. Author biography.

\end{document}